\documentclass[reprint,aps,prb,amsmath,amssymb,twocolumn,showkeys, superscriptaddress,floatfix,showkeys]{revtex4-2}
\usepackage{graphicx}
\usepackage{dcolumn}
\usepackage{bm}
\usepackage{bbm} 
\usepackage[colorlinks = true, allcolors = blue]{hyperref}
\usepackage{verbatim}
\usepackage{soul}
\usepackage{float}
\usepackage{xcolor}
\usepackage{multirow}
\usepackage{tikz}
\usetikzlibrary{shapes.geometric, arrows, shadows,positioning}

\begin{document}
\title{Phonon-assisted Casimir interactions between piezoelectric materials}

\author{Dai-Nam Le}
\email{dainamle@usf.edu}
\affiliation{Department of Physics, University of South Florida, Tampa, Florida 33620, USA}

\author{Pablo Rodriguez-Lopez}
\email{pablo.ropez@urjc.es}
\affiliation{{\'A}rea de Electromagnetismo and Grupo Interdisciplinar de Sistemas Complejos (GISC), Universidad Rey Juan Carlos, 28933, M{\'o}stoles, Madrid, Spain}

\author{Lilia M. Woods} 
\email{lmwoods@usf.edu}
\thanks{Author to whom all correspondence should be addressed}
\affiliation{Department of Physics, University of South Florida, Tampa, Florida 33620, USA}

\date{\today}

\begin{abstract}

The strong coupling between electromagnetic field and lattice oscillation in piezoelectric materials gives rise to phonon polariton excitations. Such quasiparticles open up new directions in modulating the ubiquitous Casimir force. Here by utilizing the generalized Born-Huang hydrodynamics model, three types of phonons in piezoelectrics are studied: longitudinal optical phonon, transverse optical phonon and phonon polariton. The phonon-electromagnetic coupling results in a complex set of Fresnel reflection matrices which prevents the utilization of the standard Lifshitz approach for calculating Casimir forces in the imaginary frequency domain. Our calculations are based on an approach within real frequency and finite temperatures, through which various regimes of the Casimir interaction are examined. Our study shows that piezoelectrics emerge as a set of materials where this ubiquitous force can be controlled via phonon properties for the first time. The Casimir interaction appears as a suitable means to distinguish between different types of surface phonon polaritons associated with different structural piezoelectric polytypes.

\end{abstract}
\maketitle

\section{\label{sec:1}Introduction}
Electromagnetic fluctuations between objects separated by a gap and the exchange of spatially mismatch of modes between them give rise to a ubiquitous interaction. This type of force was  orginally considered for perfect metals by H. Casimir \cite{casimir1948}. Since then, many authors have shown that the Casimir interaction can serve as a probing platform for the basic physics of the materials making up the objects and their geometry \cite{Woods2016,Klimchitskaya2009}. Experimental advances have demonstrated that Casimir forces play an important role in quantum levitation and trapping, mechanical actuators and parametric amplifiers, and self-assembly processes among others \cite{science2001, Woods2012, science2019, nature2021,Sedighi2015}. 

Although the Casimir force is universally present between objects, its scaling law, magnitude, and even sign are strongly dependent on the properties of the interacting materials. Particularly, the optical response, which of course is closely related to the electronic structure, plays a crucial role for specific features in the Casimir force. For example, topologically nontrivial properties in Dirac materials, Weyl semimetals, Chern insulators, twisted bilayer graphenes at magic angles result in unique fingerprints in Casimir phenomena \cite{Pablo2014, Woods2017, Pablo2020, Pablo2023}. Also, objects with reduced dimensionality experience much enhanced thermal fluctuation effects in the interaction, while strong anisotropy can even result in Casimir torque \cite{Le2022,Pablo2024}.

New materials have stimulated research in the field of Casimir interactions, and much of the work has been always associated with in-depth studies of the optical response properties as dictated by the underlying electronic structure of the systems. {\it Phonons}, however, are rarely discussed in the context of Casimir physics. This is typically justified as one expects negligible phonon participation in the electromagnetic exchange at separations exceeding the very short phonon mean free path due to the weak phonon-photon coupling. Nevertheless, in polar materials the hybridization between transverse optical phonon modes and photon excitations lead to tunable phonon polariton resonances in the dielectric function, which can affect the magnitude of the Casimir energy \cite{Sedighi2014,Babamahdi2019}. 

A subclass of polar materials composed of {\it piezoelectrics} is characterized by strong coupling between the macroscopic electric field $\mathbf{E}$ and the mechanical deformation vector $\mathbf{u}$ induced by phonons \cite{Ordonez2014,Auldbook}. The arising surface phonon polaritons (PhPs) create additional channels of fluctuation induced interactions between piezoelectric materials. Depending on the surface of the material, as is the case for different SiC polytypes, hybrid longitudinal-transverse modes due to additional Bragg peaks are also possible. These hybrid modes are especially pronounced in the electromagnetic coupling between substrates with finite width potentially leading to changes in the scaling law, magnitude, and even repulsion in the force. Covering the substrates with two-dimensional layers, such as graphene for example, may lead to further modifications in the Casimir interaction.
 
In this paper, we investigate piezoelectric materials as a materials platform that can support surface PhP modes to study phonon-modulated Casimir interactions for the first time. Phonon polariton modes are captured explicitly in the coupled elastic and electromagnetic boundary conditions, which are then taken in the Casimir force calculations within a real frequency formalism. Hybrid longitudinal-transverse excitations result in unexpected functionalities especially for substrates with finite thickness due to unusual interplay between quantum and thermal effects. Graphene monolayers covering the surfaces of the piezoelectric objects result in much enhanced thermal fluctuations which ultimately diminish the manifestation of piezoelectricity in the Casimir force.

\section{\label{sec:2}Electromagnetic Fields and Phonon Polaritons in Piezoelectrics}

\begin{figure}[htbp]
\begin{center}
\includegraphics[width = 0.9 \columnwidth]{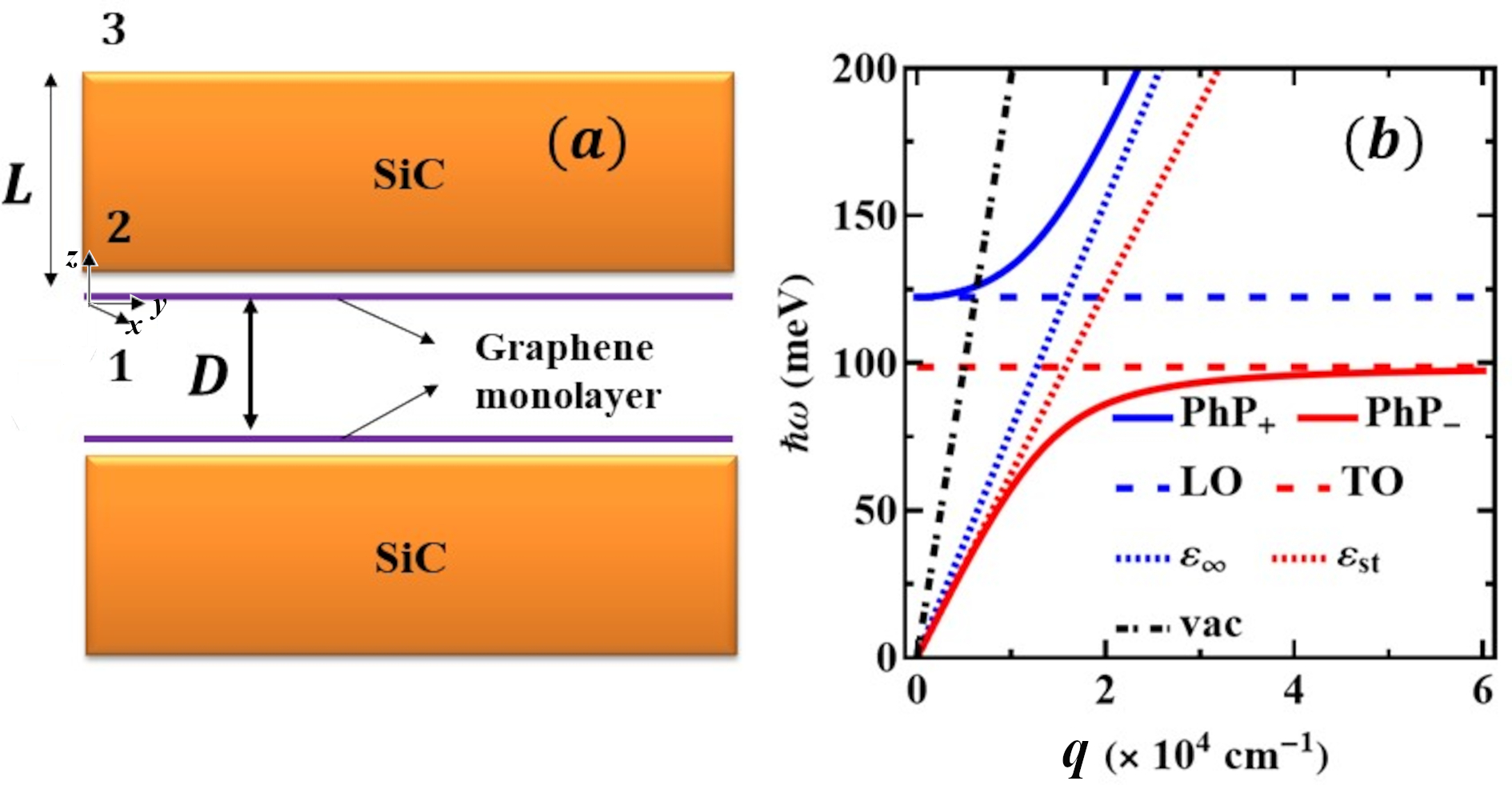}
\caption{\label{fig:1}(a) Schematics of the considered system consisting of two  graphene-coated SiC films with thickness $L$ separated by a distance $d$. (b) Phonon dispersions in bulk SiC with PhP branches together with LO and TO modes.  The light cone in vacuum (black dotted) and in a dielectric with $\varepsilon_{\infty}$ (blue dotted), $\varepsilon_{st}$ (red dotted) are also given as references. }
\end{center}
\end{figure}

The distinct feature in a piezoelectric material is that an acoustic wave impinging on its surface extends outside of the material as a decaying evanescent electric field giving rise to PhP modes \cite{Auldbook,Ridley1993, Ridley1994}. Bringing another piezoelectric nearby enables the exchange of fluctuating modes in the separating gap  creating conditions for phonon-modulated interactions. The system considered here is schematically given in Fig. \ref{fig:1} showing two identical planar piezoelectric objects with finite thickness $L$ and separated by a distance $D$. Two-dimensional layers, such as graphene, may also cover the piezoelctric surfaces forming the gap region between them. 

Let is first consider the phonon and electromagnetic modes and their hybridization in a single material. For this purpose, we utilize the generalized Born-Huang hydrodynamic model \cite{Babiker1986, Ridley1993, Ridley1994, Gubbin2019, Gubbin2020}, in which the time evolution of the ions relative mechanical displacement $\mathbf{u}$ is given as a driven harmonic oscillator governed by the electric field $\mathbf{E}$, 
\begin{eqnarray}
    \label{eqn:u-1}
&& \rho_i \left[ \frac{\partial^2}{\partial t^2} + \gamma \frac{\partial}{\partial t} + \omega_{TO}^2 \right] \mathbf{u} = e_i \mathbf{E} + \nonumber\\
&& \quad \quad \quad \quad \quad + \rho_i \left[  \beta_{TO}^2 \pmb{\nabla} \times \left( \pmb{\nabla} \times \mathbf{u} \right) - \beta_{LO}^2 \pmb{\nabla} (\pmb{\nabla}\cdot \mathbf{u})\right],
\end{eqnarray}
where $\rho_{i}$ and $e_i$ are the ionic mass and charge densities, respectively, and $\gamma$ is a damping constant. The speeds of the transverse optical (TO) and longitudinal optical (LO)  phonon are denoted as $\beta_{TO}$ and $\beta_{LO}$, respectively, and $\omega_{TO}$ is the TO phonon frequency at the center of the Brillouin zone of the material. The relation $\omega_{LO}^2 - \omega_{TO}^2 = e_i^2/\varepsilon_{\infty} \varepsilon_{0} \rho_i$ with $\varepsilon_{\infty}$ being the high frequency relative permitivity of the material ($\varepsilon_{0}$ - vacuum permittivity) establishes an inter-dependence between the characteristic LO and TO phonon frequencies \cite{Ridley1993, Ridley1994, Gubbin2020}. 

The coupling between the electric field and mechanical wave propagation is further expressed in the electrical polarization $\mathbf{P} = \left( \varepsilon_{\infty} - 1 \right) \varepsilon_0 \mathbf{E} + e_i \mathbf{u}$ of the piezoelctric. As a result, the Gauss' law $\pmb{\nabla} \cdot \mathbf{D} = 0$ is also modified since the displacement electric field now not only depends on electric field, but also on mechanical deformation $\mathbf{D} = \varepsilon_0 \mathbf{E} + \mathbf{P} = \varepsilon_{\infty} \varepsilon_0 \mathbf{E} + e_i \mathbf{u}$. 

The mechanical wave propagation in $\mathbf{P}$ together with the electromagnetic fields from the Maxwell's equations must be supplemented by appropriate boundary conditions at the piezoelectric surface. The generalized Born-Huang model distinguishes three types of phonon modes for the ionic vibrations \cite{Iadonisi2008,Lee1997}. One type are the LO phonons characterized by longitudinal displacement that satisfies $\pmb{\nabla} \times \mathbf{u}_{LO} = 0$. Another type are the TO phonons obeying $\pmb{\nabla} \cdot \mathbf{u}_{TO} = 0$ and they are a purely mechanical wave which does not induce any electromagnetic field in the material. The third type are the PhP modes associated with vibrations confined at the surface of the material whose displacement vector satisfies $\pmb{\nabla} \cdot \mathbf{u}_{PhP} = 0$. The characteristic properties of the LO, TO, and PhP modes in terms of their dispersion relations, effective dielectric functions, and corresponding electroamgnetic fields according to the Born-Huang model are summarized in Table \ref{tab:1}. 

To better understand the different excitations in a bulk piezoelectric, we consider 2H-SiC as a representative material. In Fig. \ref{fig:1}(b) we show the dispersion relations of LO, TO and PhP excitations using the following parameters \cite{Gubbin2019}: $\omega_{TO} = 795 \text{ cm}^{-1}$, $\varepsilon_{\infty} = 6.5$, $\varepsilon_{st} = 10.0$, $\beta_{TO} = 9.15 \times 10^5 \text{ cm}/\text{s}$, $\beta_{LO} = 15.39 \times 10^5 \text{ cm}/\text{s}$, $\gamma = 4 \text{ cm}^{-1}$ while the LO phonon frequency is determined from Lyddane--Sachs--Teller relation $\omega_{LO} = \omega_{TO} \sqrt{\varepsilon_{st} / \varepsilon_{\infty}}$. There are two PhP branches and they both are of transverse polarization due to hybridized TO phonons and photons. The propagating PhP$_+$ branch is optically active for  wave vectors $q < 0.65 \times 10^7 \text{ cm}^{-1}$, while the decaying evanescent PhP$_-$ branch is optically active for all $k$. 

\onecolumngrid

\begin{table}[H]
\caption{\label{tab:1}Dispersion relations, effective dielectric functions, and electromagnetic fields of the LO, TO and PhP modes in the generalized Born-Huang model \cite{Ridley1993, Ridley1994}. Here 
$\mathbf{q}=(\mathbf{q}_{\parallel},q_z)$, such that $\mathbf{q}_{\parallel}=(q_x,q_y)$ and $q^2=q_{\parallel}^2+q_z^2$.}
\begin{ruledtabular}
\begin{tabular}{llll}
Mode  & LO & TO & PhP \\
\hline\\
Dispersion & $\omega^2 = \omega_{LO}^2 - \beta_{LO}^2 q^2$ & $\omega^2 = \omega_{TO}^2 - \beta_{TO}^2 q^2$  & $\omega^2 = q^2 c^2/{\varepsilon_{PhP} (\omega, q)}$ \\
Dielectric function & $\varepsilon_{LO} (\omega, q) = 0$ & $\varepsilon_{TO} (\omega, q) \to \infty$ & $\frac{\varepsilon_{PhP} (\omega, q)}{\varepsilon_{\infty}} = \frac{\omega_{LO}^2 - \omega (\omega + i \gamma) - \beta_{TO}^2 q^2}{\omega_{TO}^2 - \omega (\omega + i \gamma) - \beta_{TO}^2 q^2} $ \\
EM field & $\mathbf{E}_{LO} = - \dfrac{e_i}{\varepsilon_{\infty} \varepsilon_0} \mathbf{u}_{LO}$ & $\mathbf{E}_{TO} = 0$ & $\mathbf{E}_{PhP} = \frac{e_i}{\left[ \varepsilon_{PhP} (\omega, q) -\varepsilon_{\infty} \right] \varepsilon_0} \mathbf{u}_{PhP} $ \\
& $\mathbf{B}_{LO} = 0$ & $\mathbf{B}_{TO} = 0$ & $\mathbf{B}_{PhP} = \frac{e_i}{\left[ \varepsilon_{PhP} (\omega, q) -\varepsilon_{\infty} \right] \varepsilon_0 \omega} \mathbf{q} \times \mathbf{u}_{PhP} $\\
\end{tabular}
\end{ruledtabular}
\end{table}
\twocolumngrid

We now continue with examining the boundary conditions at the surface of the piezoelectric material. The electromagnetic boundary conditions are consistent with Maxwell's equations and they include continuity of $D_{z}$, $E_{x}$, $E_{y}$, $B_{z}$, $H_{x}$, $H_{y}$ components at the surface $z=0$. Additionally, the mechanical ionic displacement satisfies the elastic boundary conditions of $\mathbf{u} = (u_x, u_y, u_z)$ being continuous at $z=0$. Since vacuum has no ionic displacement, then $\mathbf{u}_{LO} + \mathbf{u}_{TO}+\mathbf{u}_{PhP} = \pmb{0}$ at $z=0$. The resolution of both types of conditions requires all three types of phonons, LO, TO, and PhP excitations, which leads to complex inter-relations for the reflection coefficients. For the system with isotropic materials in Fig. \ref{fig:1}a, the Fresnel scattering matrix \cite{Woods2016,Klimchitskaya2009} has a diagonal form $\textbf{diag} \left\{ r_{ss} (\mathbf{q}_{\parallel}, \omega), r_{pp} (\mathbf{q}_{\parallel}, \omega) \right\} $, with coefficients for $\alpha=(s,p)$ polarization found as 
\begin{widetext}
\begin{eqnarray}
    \label{eqn:r-finite}
r_{\alpha\alpha} (\mathbf{q}_{\parallel},\omega) &=& r^{12}_{\alpha\alpha} (\mathbf{q}_{\parallel},\omega) + \frac{t^{12}_{\alpha\alpha}(\mathbf{q}_{\parallel},\omega) r^{23}_{\alpha\alpha}(\mathbf{q}_{\parallel},\omega) t^{21}_{\alpha\alpha}(\mathbf{q}_{\parallel},\omega) e^{2 i q_{PhP,z} L}}{1 - r^{21}_{\alpha\alpha}(\mathbf{q}_{\parallel},\omega) r^{23}_{\alpha\alpha}(\mathbf{q}_{\parallel},\omega) e^{2 i q_{PhP,z} L}},\\
    \label{eqn:r-ss-final}
   \lim\limits_{L\to\infty} r_{ss} \left( \mathbf{q}_{\parallel},\omega \right) &=& \frac{q_{z} - q_{PhP,z} - \mu_0 \omega \sigma_s (\mathbf{q}_{\parallel},\omega)}{q_{z} + q_{PhP,z} + \mu_0 \omega \sigma_s (\mathbf{q}_{\parallel},\omega)} \\
    \label{eqn:r-pp-final}
\lim\limits_{L\to\infty} r_{pp} \left( \mathbf{q}_{\parallel},\omega \right) &=& \frac{\varepsilon_{PhP}( \omega)q_{z} - \left( q_{PhP,z} + \Omega(\mathbf{q}_{\parallel},\omega) \right) + \mu_0 c \sigma_s (\mathbf{q}_{\parallel},\omega) \left(\frac{q_{z} c}{\omega}\right) \left( q_{PhP,z} + \Omega(\mathbf{q}_{\parallel},\omega) \right) }{ \varepsilon_{PhP}( \omega)q_{z} + \left( q_{PhP,z} + \Omega(\mathbf{q}_{\parallel},\omega) \right) + \mu_0 c \sigma_s (\mathbf{q}_{\parallel},\omega) \left(\frac{q_{z} c}{\omega}\right) \left( q_{PhP,z} + \Omega(\mathbf{q}_{\parallel},\omega) \right) },
\end{eqnarray}

\end{widetext}
where $\mathbf{q}_{\parallel}=(q_x,q_y)$ is the wave vector along the surface, $ q_{z}  = \sqrt{\omega^2/c^2 - q_{\parallel}^2}$ and $q_{PhP,z} = \sqrt{ \varepsilon_{PhP} \left( \omega, q \right) \omega^2/c^2 - q_{\parallel}^2}$. The indices $i,j = 1,2,3$ denote the regions shown in Fig. \ref{fig:1}a, while $r^{ij}$, $t^{ij}$ correspond to the reflection and transmission coefficients at the interface between the two media. For semi-infinite plates for which $L \to \infty$, Eq. \eqref{eqn:r-finite} becomes Eqs. \eqref{eqn:r-ss-final}, \eqref{eqn:r-pp-final}. Explicit expressions for $r^{ij}_{\alpha \alpha}$ and $t^{ij}_{\alpha \alpha}$ and details of their derivations are given in Sections I and II in the Supplementary Information \cite{Supp}. The atomic layers in the gap region are taken into account via their surface conductivity, which in the case of graphene is taken to be its universal value $\sigma_s(\mathbf{q}_{\parallel},\omega)= e^2 / 4 \hbar$ \cite{Falkovsky2007}.

A distinct feature arising from the boundary conditions is the  surface PhP (SPhP) mode. Such excitations are a direct consequence of the coupled electromagnetic-elastic boundary conditions, such that 
\begin{equation}
\label{eqn:Omega-LO}
\Omega(\mathbf{q}_{\parallel}, \omega) = \frac{q_{\parallel}^2 \left( q_{PhP,z} - q_{TO,z} \right)}{ q_{\parallel}^2+ q_{TO,z} q_{LO,z} } \left( \varepsilon_{PhP}(\omega)  - \varepsilon_{\infty} \right) ,
\end{equation}
in which $q_{TO,z} = \sqrt{\omega_{TO}^2 - \omega^2}/\beta_{TO}$ and $q_{LO,z} = \sqrt{\omega_{LO}^2 - \omega^2}/\beta_{TO}$ are the local out-of-plane components of the wave vectors of the TO and LO phonons. 
Eqs. \eqref{eqn:r-finite}-\eqref{eqn:r-pp-final} show that, in fact, the elastic boundary conditions affect only the $p$-polarized modes, while the $s$-polarization remains the same as from standard electromagnetic boundary conditions. The factor $\left( \varepsilon_{PhP}(\omega)  - \varepsilon_{\infty} \right)$ indicates that the electromagnetic-elastic boundary conditions term $\Omega(\mathbf{q}_{\parallel}, \omega)$ is significant especially at small frequency. Consequently, the 
dielectric function for the piezoelectric material is determined from solving the PhP excitations dispersion $\omega^2 = q^2 c^2/{\varepsilon_{PhP} (\omega, q)}$ which takes into account the nonlocal dielectric function of PhP mode in the Born-Huang model given in Table \ref{tab:1}. Since $\beta_{TO}, \beta_{LO} \ll c$, the PhP dielectric function is found as
\begin{eqnarray}
    \label{eqn:eps-PhP-final}
\frac{\varepsilon_{PhP} (\omega)}{\varepsilon_{\infty}} \approx \frac{\omega_{LO}^2 - \omega (\omega+i \gamma) -  \frac{\varepsilon_{\infty} \omega^2 \beta_{TO}^2}{c^2} \frac{\omega_{LO}^2 - \omega (\omega+i \gamma)}{ \omega_{TO}^2 - \omega (\omega+i \gamma) }}{ \omega_{TO}^2 - \omega (\omega+i \gamma) - \frac{\varepsilon_{\infty}\omega^2  \beta_{TO}^2}{c^2} \frac{\omega_{LO}^2 - \omega (\omega+i \gamma)}{ \omega_{TO}^2 - \omega (\omega+i \gamma) }}.
\end{eqnarray}

The Born-Huang hydrodynamics model is based on a continuum approximation and it has been successfully applied in a variery of piezoelectric materials  \cite{Gubbin2019,Xu2023,Xie2023,Gubin2020,Tranchant2019}. SiC, the material of interest here, has several polytypes with distinct electromagnetic spectra. Specifically, for polytypes, such as 4H-, 6H-, 8H-, 10H-SiC, an elongated $c$-axis in the lattice results in additional Bragg peaks in the spectrum causing {\it zone folding} of the LO phonon dispersion $\omega (\mathbf{q})$ in the center of the Brillouin zone, such that $q_{\parallel} \to q_{\parallel} + q_{m,n}$ where $q_{m,n} = \frac{2 m \pi}{n c_2}$ and $m = 1,2,\ldots, n-1$ is the number of Bragg peak scattering. Signatures of the zone folding lying in the Reststrahlen band between TO and LO phonon frequencies have been observed experimentally in the Raman spectra of various SiC polytypes \cite{Feldman1968, Nakashima2001, Gubbin2019,Caldwell2013}. For a 4H-SiC polytype, considered here, zone folding is accounted by taking 
$q_{\parallel}^2\rightarrow q_{\parallel}(q_{\parallel}+q_{m,n})$ in Eq. \eqref{eqn:Omega-LO}, where $m = 1, n = 2$ and $q_{1,2} = \pi / c_2$ with $c_2 = 5 \text{ \AA}$ \cite{Gubbin2019} (Details in Section IA in the Supplementary Information \cite{Supp}).

\onecolumngrid

\begin{figure}[htpb]
\begin{center}
\includegraphics[width = 0.96 \columnwidth]{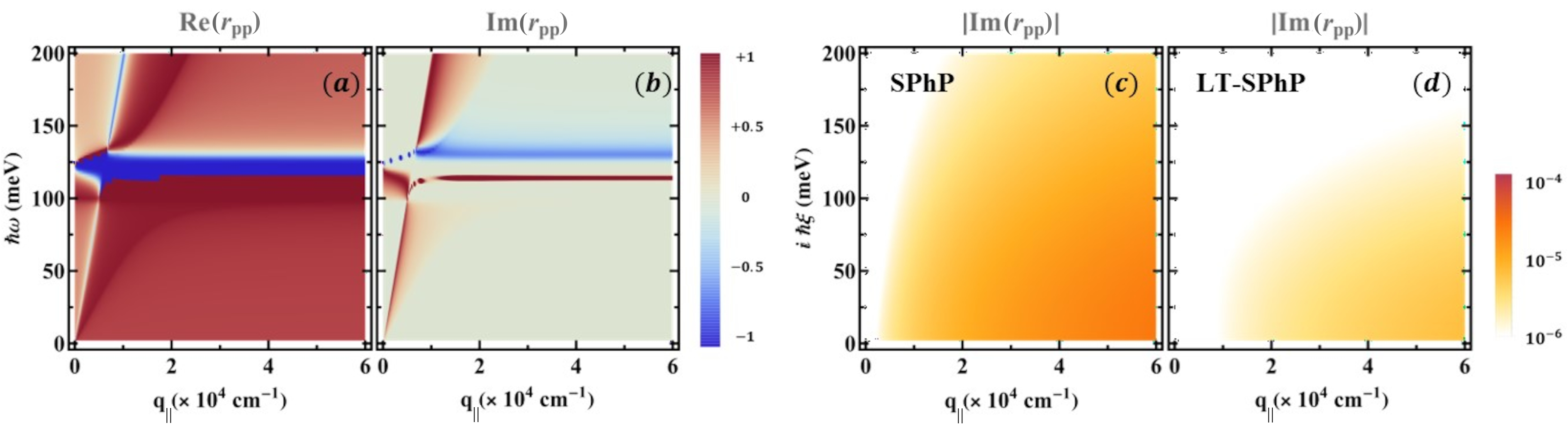}
\caption{\label{fig:2} Density plots in the $(\hbar \omega , q)$ domain of a semi-infinite SiC substrate showing  different PhP, LO, and TO branches and their dispersions in: (a) $\textbf{Re} (r_{pp})$ and (b) $\textbf{Im} (r_{pp})$. Density plots in the $(i\hbar \xi, q)$ domain of a semi-infinite SiC substrate showing $|\textbf{Im} (r_{pp})|$ of a SiC substrate supporting: (c) SPhP and (d) LT-SPhP.}
\end{center}
\end{figure}

\twocolumngrid 

In Figs. \ref{fig:2} (a, b), the density plots of the real and imaginary parts of $r_{pp}$ for a semi-infinite 4H-SiC plate are given. We find that in addition to the PhP$_{+}$ and PhP$_{-}$ branches (broad dark red regions in Fig.  \ref{fig:2}(a)), rather localized modes close to the photon line and between $\omega_{TO}, \omega_{LO}$ (blue regions) appear. These new modes are a direct consequence of the combined effect from the zone-folded phonons and SPhPs and they are associated with $\varepsilon_{PhP} (\omega) q_{z} + q_{PhP,z} + \Omega (\mathbf{q}_{\parallel},\omega) = 0$. One obtains that the regions near the surface modes SPhP${}_{+}$ and SPhP${}_{-}$ branches are characterized by $ \textbf{Re }(r_{pp}) \to \pm \infty$. On the other hand, $ \textbf {Im }(r_{pp}) \to \infty $ is mostly localized along SPhP${}_{-}$ and $ \textbf {Im }(r_{pp}) \to - \infty$ along SPhP${}_{+}$ near LO line. Anti-crossing due to the strong interaction between zone-folded LO phonons and SPhPs is observed in Fig. \ref{fig:2}(b). In fact, the resultant quasiparticle, longitudinal-transverse SPhP (LT-SPhP), contains an admixture of longitudinal and transverse polarization types. LT-SPhPs are accessible for a wide range of wave vectors and they have recently been observed in SiC systems \cite{Gubbin2019}.

\section{\label{sec:3}Casimir force in real frequency for planar objects}

Lifshitz formalism \cite{Lifshitz1956} is widely used for obtaining interaction energies and forces between objects. Calculations are typically done using  summation over the imaginary Matsubara frequency domain $\omega = i\xi$, where dielectric functions are monotonically decreasing functions. However, the reflection coefficient $r_{pp} \left(\mathbf{q}_{\parallel}, i \xi \right)$ is a complex quantity. For plates supporting only SPhPs (Fig. \ref{fig:2}(c)) and plates with LT-SPhPs excitations (Fig. \ref{fig:2}(d)) $r_{pp}$ has complex nature, which is more pronounced especially for larger wave vectors. This implies that  one must consider Rytov's theory of fluctuating electrodynamics in real frequencies \cite{Rytovbook}, as originally proposed by Lifshitz \cite{Lifshitz1956}. Particularly, the pressure between two identical planar objects \cite{Lifshitz1956, Antezza2006, Svetovoy2007, Bordag2009}  is found as
\begin{widetext} 
\begin{eqnarray}
    \label{eqn:Casimir-force}
P(D,T) = &&\sum\limits_{\alpha = s, p} \left[ P_{\text{prop}}^{\alpha} (D,T) +  P_{\text{evan}}^{\alpha} (D,T) \right], \\
    \label{eqn:Casimir-prop}
     P_{\text{prop}}^{\alpha} (D,T) =  && \frac{\hbar}{4\pi^3} \textbf{Re}   \int\limits_0^{+\infty} d \omega \; \eta (\omega,T) \iint\limits_{q < \omega/c} d^2 \mathbf{q}_{\parallel}  \frac{ q_{z} r_{\alpha\alpha}^{(1)} (\mathbf{q}_{\parallel},\omega) r_{\alpha\alpha}^{(2)} (\mathbf{q}_{\parallel},\omega) e^{2 i q_{z} D}  }{1 - r_{\alpha\alpha}^{(1)} (\mathbf{q}_{\parallel},\omega) r_{\alpha\alpha}^{(2)} (\mathbf{q}_{\parallel},\omega) e^{2 i q_{z} D}},\\
    \label{eqn:Casimir-evan}
     P_{\text{evan}}^{\alpha} (D,T) =  &&  \frac{\hbar}{4\pi^3}\textbf{Re}   \int\limits_0^{+\infty} d \omega \; \eta (\omega,T) \iint\limits_{q > \omega/c} d^2 \mathbf{q}_{\parallel}  \frac{ q_{z} r_{\alpha\alpha}^{(1)} (\mathbf{q}_{\parallel},\omega) r_{\alpha\alpha}^{(2)} (\mathbf{q}_{\parallel},\omega) e^{2 i q_{z} D}  }{1 - r_{\alpha\alpha}^{(1)} (\mathbf{q}_{\parallel},\omega) r_{\alpha\alpha}^{(2)} (\mathbf{q}_{\parallel},\omega) e^{2 i q_{z} D}},
\end{eqnarray}
\end{widetext}
where $\eta (\omega, T) =  \coth \left( \frac{\hbar \omega}{2 k_B T}\right)$ is the Planck distribution function at temperature $T$. The quantum mechanical limit is found by taking $\eta (\omega, 0) \to 1$, while the thermal limit corresponds to $\eta (\omega, T) \to 2k_B T/\hbar \omega$. 

\onecolumngrid

\begin{figure}[htbp]
\begin{center}
\includegraphics[width = 0.95 \columnwidth]{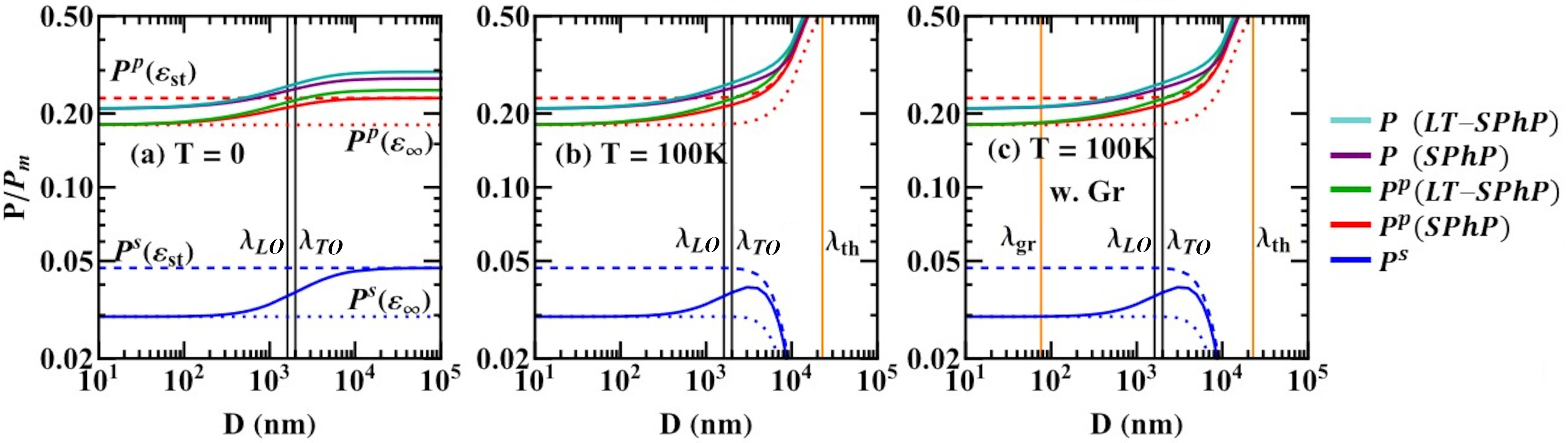}
\caption{\label{fig:3}Casimir pressure $P$ normalized by the perfect metal limit $P_m = - \frac{\pi^2 \hbar c}{240 D^4}$ between: (a) two semi-infinite SiC plates at $T=0$ K; (b) two semi-infinite SiC plates at $T=100$ K; (c) two semi-infinite SiC plates covered by graphene sheets at $T=100$ K. The characteristic wavelengths are denoted as 
$\lambda_{TO}=\frac{c}{\omega_{TO}}, \lambda_{LO}=\frac{c}{\omega_{LO}}, \lambda_{th}=\frac{\hbar c}{k_BT}, \lambda_{gr}=\frac{\hbar v_F}{k_BT}$ where $\omega_{TO} = 795 \text{ cm}^{-1}$, $\omega_{LO} = 986 \text{ cm}^{-1}$, $v_F=10^5$ m/s. The Casimir pressures for semi-infinite plates with $\varepsilon_{\infty} = 6.5$ and $\varepsilon_{st} = 10.0$ (characteristic dielectric constants for SiC) are also shown.}
\end{center}
\end{figure}
\twocolumngrid

Applying Wick rotation $\omega \to i \xi$ to $P^{\alpha}(D,T) =  P_{\text{prop}}^{\alpha} (D,T) +  P_{\text{evan}}^{\alpha} (D,T)$ and combining with the fact that the Matsubara frequencies $\xi_n=2\pi n k_BT/\hbar$ appear as poles of $ \eta (i \xi, T) $, the Casimir pressure is found as
\begin{widetext}
\begin{eqnarray}
    \label{eqn:Casimir-force-imag-real}
&&P^{\alpha}(D,T) =  P_{\text{prop}}^{\alpha} (D,T) +  P_{\text{evan}}^{\alpha} (D,T) = - \frac{k_B T}{2 \pi^2} \textbf{Re } {\sum_{n=0}^{\infty}}^{\prime} \int\limits_0^{+\infty} d^2 \mathbf{q}_{\parallel} \left\{ \kappa_z  \frac{ r_{\alpha\alpha}^{(1)}(\mathbf{q}_{\parallel}, i\xi_n) r_{\alpha\alpha}^{(2)}(\mathbf{q}_{\parallel}, i\xi_n) e^{- 2 \kappa_z D}}{1 -  r_{\alpha\alpha}^{(1)}(\mathbf{q}_{\parallel}, i\xi_n) r_{\alpha\alpha}^{(2)}(\mathbf{q}_{\parallel}, i\xi_n) e^{- 2 \kappa_z D}}  \right\}.
\end{eqnarray}

\end{widetext}
where $\kappa_z=\sqrt{\xi_n^2/c^2 + q_{\parallel}^2}$. The prime in the summation sign indicates that the $n=0$ Matsubara term is weighted by $1/2$. 
The above equation can now be applied to obtain the interaction pressure between the piezoelectric materials by taking into account the complex nature of their reflection coefficients.

\section{\label{sec:4}Casimir force between piezoelectric materials}

\subsection{Semi-infinite plates}

We first consider the case of semi-infinite plates in Fig. \ref{fig:1}(a)) with corresponding Fresnel reflection coefficients given in  Eqs. \eqref{eqn:r-ss-final}, \eqref{eqn:r-pp-final}. Results for the interaction between plates without graphene at $T=0$ K are shown in Fig. \ref{fig:3}(a). One finds that $P^s$ and $P^p$ follow similar scaling laws, however, the contribution from $P^p$ to the tolal pressure is dominant. Both $P^s$ and $P^p$ are bound by their respective limits associated with the $\varepsilon_{\infty}$ and $\varepsilon_{st}$ dielectric constants. It appears that $P^s, P^p \sim 1/D^{-4}$ everywhere except in the $0.5-10$ microns region, in which the Casimir interaction transitions between the limiting coupling found with $\varepsilon_{st}$ and $\varepsilon_{\infty}$. This transition region is comparable to the phonon wavelengths $\lambda_{LO} = \frac{c}{\omega_{LO}} = 1.6$ and $\lambda_{TO} = \frac{c}{\omega_{TO}} = 2$ microns. The calculations show that the Casimir interaction for substrates supporting SPhPs or LT-SPhPs excitations is identical at small separations, while in the sub-micron and larger region (consistent with the long-wavelength limit), $P$(LT-SPhS) is slightly larger than $P$(SPhS) (see Section III in the Supplementary Information \cite{Supp} for details).

As an example representation for the role of temperature, the Casimir pressure at $T=100$ K  is given in Fig. \ref{fig:2}(b), where the characteristic thermal wavelength $\lambda_{th}=\frac{\hbar c}{k_BT}$ are also shown. One finds that in the $(\lambda_{LO}, \lambda_{th})$ region, $P^s, P^p$ experience a scaling law transition from $1/D^4$ to $1/D^3$ signaling the onset of thermal fluctuations at larger separations. More precisely, one can find the thermal transition distance $D_{th}$ from $P(T=0) = P_{n=0}(T)$, which yields $D_{th} \approx  8$ microns at $T = 100$ K. (see Section V in the Supplementary Information \cite{Supp}). Similar to the quantum mechanical limit, the Casimir pressure is dominated by the $p$-polarized modes with little distinction between SPhPs and LT-SPhPs modes at sub-micron and larger separations.

The inclusion of graphene layers on top of the surfaces of the semi-infinite piezolectrics does not change significantly the Casimir interaction at $T=0$ K. This is evident in Fig. \ref{fig:2}(c), which also shows the greatly diminished $P^s$ role in the total pressure. One further finds that graphene pushes the quantum-to-thermal transition to lower distances. This is not surprising given the much reduced characteristic thermal wavelength of graphene $\lambda_{gr}=\frac{\hbar v_F}{k_B T}$, where $v_F=10^5$ m/s is the Fermi velocity \cite{Woods2010, Le2022, Drosdoff2012}. As a result, the $1/D^{-4}\rightarrow 1/D^{-3}$ transition happens in a broader region between $\lambda_{gr}$ and $\lambda_{th}$. From $P(T=0) = P_{n=0}(T)$ a transition distance $D_{th} \approx 5$ microns is found for $T=100$ (see Section V in the Supplementary Information \cite{Supp}). As $T$ increases, however, the $\lambda_{gr}< D_{th} < \lambda_{th}$ shrinks indicating a sharper quantum-to-thermal transition in the Casimir interaction as shown for $T=300$ K in the Supplementary Information \cite{Supp}.

\onecolumngrid

\begin{figure}[H]
\begin{center}
\includegraphics[width = 0.98\columnwidth]{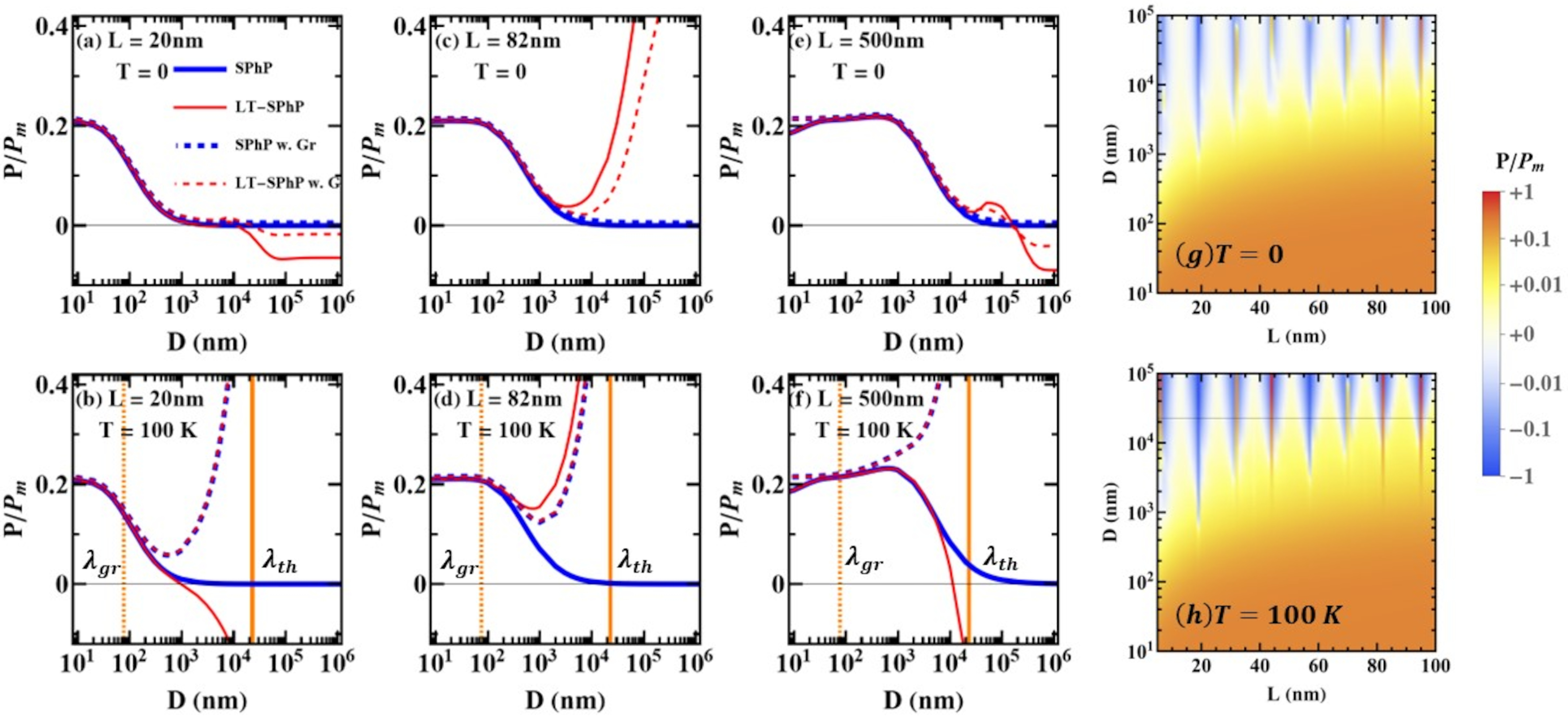}
\caption{\label{fig:4}Casimir pressure $P$ normalized by the perfect metal limit $P_m = - \frac{\pi^2 \hbar c}{240 D^4}$ between two SiC plates with: (a) thickness $L=20$ nm, $T=0$ K; (b) thickness $L=20$ nm, $T=100$ K; (c) thickness $L=82$ nm, $T=0$ K; (d) thickness $L=82$ nm, $T=100$ K; (e) thickness $L=500$ nm, $T=0$ K; (f) thickness $L=500$ nm, $T=100$ K. Results are shown for plates supporting SPhPs (solid blue), LT-SPhPs (solid red), plates supporting SPhPs and covered with graphene (dashed blue), plates supporting LT-SPhPs and covered with graphene (dashed red); (g) Density plots showing the $D$ vs $L$ dependence of the Casimir pressure $P/P_m$ between two plates supporting LT-SPhPs at $T = 100$ K.}
\end{center}
\end{figure}
\twocolumngrid

\subsection{Finite Thickness Plates}

Let us now consider the Casimir interaction between two SiC plates with a finite thickness distinguishing SPhPs vs LT-SPhPs and the role of graphene. Numerical calculations for $L = 20, 82$ and $500$ nm at $T=0$ K are given in Figs. \ref{fig:4} (a,c,e). We find that for SiC supporting both types of modes, the pressure decreases non-monotonically as $D$ increases. It appears that for $D\lesssim L$, $P$ exhibits similar behavior as the one found for semi-infinite plates (Fig. \ref{fig:3} (a)). As $D$ becomes larger, however, the $P/P_m$ experiences non-monotonical decrease with a sharper descent for thinner plates. Particularly, our calculations show that for plates with SPhPs modes $P/P_m \sim 1/D^2$, in other words, the scaling law has changed from $1/D^4$ for semi-infinite plates to $1/D^6$ upon increasing the separation distance \cite{Welsch2003, Zhao2011}. This type of scaling law is indicative that the SiC plates behave as two dielectrics. 

In fact, since $\Omega (\mathbf{q}_{\parallel},\omega)\sim q^2_{\parallel}$ (Eq. \eqref{eqn:Omega-LO}) the elastic boundary terms have no contribution to the long wavelength limit of the interaction, thus the reflection and transmission coefficients $r_{pp}^{ij}, t_{pp}^{ij}$ of p-polarized waves obeys the same constrains as dielectrics described with $\varepsilon_{st}$. Consequently, the Fresnel reflection coefficient from Eq. \ref{eqn:r-finite} is found to have the following asymptotic behavior 
$r_{\alpha \alpha}(\mathbf{q}_{\parallel},i\xi) \propto \xi L/c $ for both type of polarizations at long wavelength limit $q_{\parallel} \to 0$. It is the $r_{\alpha \alpha}\sim \xi$ relation that gives the additional powers in the separation scaling law of the Casimir pressure, such that (details in Section IVA in the Supplementary Information \cite{Supp})
\begin{eqnarray}
P (D \gg L) \approx \left\{ \begin{array}{ll}
- \frac{ (\varepsilon_{st}-1)^2 \left( 9 \varepsilon_{st}^2 + 10 \varepsilon_{st} + 4 \right) L^2\hbar c}{32 \pi^2 \varepsilon_{st}^2 D^6} & T = 0 \\
- \frac{3 (\varepsilon_{st}-1)^2 L^2 k_B T}{32 \pi D^5} & T \gg \frac{\hbar c}{D k_B}
\end{array}
 \right.
\end{eqnarray}
The inclusion of graphene has a minor effect as long as $D \lesssim L$ indicating that the quantum regime is dominated by the properties of the piezoelectric plate itself. At sub-micron and larger separations, however, the interaction is almost completely dominated by graphene $P \approx P_{gr} \approx 5.38 \times 10^{-2} P_m $ recovering the expected $1/D^4$ scaling law. 

When the SiC plates support LT-SPhPs, however, $P$ may have significant deviations from the one obtained for plates with SPhPs modes, which are especially pronounced at larger $D$. For example, the interaction becomes negative with $1/D^4$ scaling law at $D>10^4$ nm for $L=20$ nm plates and $D>10^5$ nm for $L=500$ nm plates. This repulsive effect is reduced upon inclusion of graphene layers. On the other hand, for plates with $L=82$ nm, the interaction becomes strongly attractive approaching the perfect metal limit  for $D > 10^4$ nm. Graphene layers reduce by close to a half this enhancement. Thus unlike SiC plates supporting SPhPs, the LT-SPhPs modes in the long range limit cannot be completely dominated by graphene partly due to the same $1/D^4$ scaling law.

Numerical calculations for $L = 20, 82, 500$ nm at $T=100$ K are given in Figs. \ref{fig:4} (b,d,f). The results show that temperature alters the scaling law from $1/D^6$ to $1/D^5$ for plates supporting only SPhPs. LT-SPhP modes, however, have a much pronounced effect as Fig. \ref{fig:4}(b,f) show a stronger repulsive effect arising at smaller separations (about an order smaller as compared to the $T=0$ K case) with the typical $1/D^3$ scaling law of thermal fluctuation. The transition to $1/D^3$ behavior is also enhanced when compared to plates with SPhPs, as Fig. \ref{fig:4}(d) shows a steeper scaling law change with larger magnitude of attraction. The inclusion of graphene accentuates the role of thermal fluctuations. As mentioned earlier, since graphene is responsible for almost complete reflection of the low frequency  $p$-modes, the Casimir pressure becomes dominated by thermal effects around $D \approx 5$ microns consistent with the thermal wavelength of graphene. The repulsion from LT-SPhPs is washed away rendering a very similar behavior of finite thickness piezoelectric plates with SPhPs.  

Fig. \ref{fig:4}(a-f) shows that there is an unusual interplay between $D$ and $L$ in the interaction between plates supporting LT-SPhPs. To further examine this relation, Fig. \ref{fig:4}(g,h) shows a $D$ vs $L$ density map of the Casimir pressure at $T=0$ and 100 K. The broad yellow region corresponds to $D\leq L$ showing that the interaction is similar to the one of semi-infinite plates (discussed earlier). An unusual feature is the oscillatory-like behavior outside of that limit marked by positive and negative peaks. This is unique to the LT-SPhPs excitations and it is directly linked to $\Omega_{\pm}\sim 
\frac{q_{1,2}}{q_{LO}}$ at the long wavelength limit with  $q_{LO} = \frac{\omega_{LO}}{\beta_L}$ being the LO phonon wave vector. Thus, unlike SPhPs for which $\Omega\sim q_{\parallel}^2$, the characteristics frequencies for LT-SPhPs enetring the Fresnel coefficients are constant associated with the phonon zone folding parameter $q_{1,2}=\frac{\pi}{c_2}$.  We find that 
\begin{equation}
P (D \gg L)  \approx \left\{
\begin{matrix}
- \frac{\pi^2 \hbar c}{240 D^4} \mathcal{R} (L) & T = 0 \\
- \frac{\zeta(3) k_B T}{8 \pi D^3} \mathcal{R} (L) & T \gg \frac{\hbar c}{k_B D}
\end{matrix}
 \right.,       
\end{equation}
\begin{equation}
\mathcal{R} (L)=\dfrac{1 -  \left(\frac{\pi \varepsilon_{st} q_{LO}}{2 \left( \varepsilon_{st} - \varepsilon_{\infty} \right) q_{1,2}}\right)^2 \cot^2 \left( \frac{q_{LO} L}{2} \right) }{(1+ \left(\frac{\pi \varepsilon_{st} q_{LO}}{2 \left( \varepsilon_{st} - \varepsilon_{\infty})\right) q_{1,2}}\right)^2 \cot^2 \left( \frac{q_{LO} L}{2} \right))^2}
\end{equation}
Apparently, confining the LO phonon into the cavity results in resonant-like trapped LT-SPhPs excitations controlled by the thickness and associated with the periodic  $\cot \left( \frac{q_{LO} L}{2} \right)$. The attractive ``resonances'' occur when $\cot \left( \frac{q_{LO} L}{2} \right) = 0$, while repulsive ``resonances'' happen  for $\cot \left( \frac{q_{LO} L}{2} \right) = \pm \frac{2 \sqrt{3} \left( \varepsilon_{st} - \varepsilon_{\infty} \right) q_{1,2}}{\pi \varepsilon_{st} q_{LO}}$ (see Section IVB in the Supplementary Information  \cite{Supp}). These resonant-like features are controlled by the piezoelectric properties and the finite width of the plate through the factor $\mathcal{R} (L)$, and they appear broadened as temperature is increased. At the same time, the characteristic scaling laws at the quantum and thermal limit are preserved. The numerical simulations shown in Fig. \ref{fig:4}(g, h) are consistent with the analytical expressions at the long wavelength limit in the above equations. 

\section{\label{sec:5}Conclusions}

The ubiquitous Casimir force is typically considered as a macroscopic manifestation of quantum vacuum excitations, where phonons are not relevant. However, in piezoelectric materials, the phonon excitations can lead to surface hybrid modes which can significantly modify the force. These new developments are made possible by employing a generalized Born-Huang model. Taking SiC plate as a representative example, the model describes SPhPs, decaying quasipartices with longitudinal polarization.  The Born-Huang model can also accommodate LO phonon zone-folding in SiC polytypes, whose strong interaction with SPhPs results in a hybrid quasiparticle, LT-SPhPs, which is of {\it mixed} longitudinal and transverse polarization. 

A similar type of polarization hybridization also occurs in plasmonic systems due to the finite thickness of the substrate and the electronic pressure \cite{Ciraci2013,Newman2015,Gong2023}. In such systems, the hybrid modes are possible at a finite wave vector range. In plasmonic materials, one has to resort to ultrathin films (on the order of a few nanometers) where the strong localization of the electric field leads to hybrid plasmon polarizations characterized by a vanishing permittivity at a small frequency range \cite{Ciraci2013,Newman2015}. For such systems, the Casimir force exhibits unusual scaling as a function of the film thickness and distance separation. 

In the case of piezoelectrics, however, the hybrid phonon modes are not restricted to a specific wave vector range. One does not have to utilize ultrathin plates to observe SPhPs signatures in the Casimir interaction. Because of the SPhP excitations, the effective dielectric constants of piezoelectrics are different for low and high frequency modes. Consequently, the Casimir pressure exhibits a transition at short and long distance limits with associated transition distance around the phonon wavelengths $\lambda_{TO, LO}$. However, since SPhPs and LT-SPhPs modes behave differently in the long-wavelength limit,  the Casimir force acquires specific to each type of excitation features.

Taking into account the finite thickness of the piezoelectric plates, this distinction is even more significant. Indeed, when the separation distance is larger than the thickness of the piezoelectric plates $D \gg L$, their Casimir pressure assisted by SPhPs decays as $1/D^6$ at larger separations, similar to substrates made of typical dielectrics \cite{Welsch2003, Zhao2011}. This behavior is primarily associated with the SiC dielectric function approaching the $\varepsilon_{st}$ limit and the small long wavelength contributions of SPhS modes. The force between plates with LT-SPhPs decays as $1/D^4$ in the long range limit. It appears that in this case the hybrid modes, however, the confined LO phonons in the cavity can result in resonant-like conditions associated with the periodic $\tan \left( q_{LO} L \right)$. Positive or negative ``resonances'' in the Casimir pressure are controlled by the thickness and the optical properties of the piezoelectric material, as discussed previously.

Thermal fluctuation at nonzero temperatures dominate the quantum fluctuations as long as the separation distance is larger than thermal wavelength $D \gg \lambda_{th}$. Thermal effects change the scaling law of the quantum Casimir pressure between finite thickness plates supporting SPhP modes from $1/D^6$ to $1/D^5$ for SPhP modes, while $1/D^4$ to $1/D^3$ transition is found for finite thickness plates supporting LT-SPhP modes. Similar to zero temperature, confining LO phonon into cavity modes affects on LT-SPhP modes and makes the force be repulsive at appropriate thickness.

Coating the piezoelectric plates by graphene monolayers also has interesting consequences in the Casimir interaction. Due to its constant optical conductivity at small frequencies, there is an almost complete reflection of the low frequency $p$-modes. As a result, the effects from SPhPs and LT-SPhPs excitations at finite temperature are significantly reduced now  in the Casimir force for separations $D > \lambda_{th}$. Graphene, however, is almost inconsequential for smaller separations regardless of temperature. 

Our study suggests that phonons do have strong effects on the Casimir force whose "typical" source are electromagnetic fluctuations between objects. Piezoelectric materials appear to be a suitable platform for observing consequences of surface plasmon polaritons through the exchange of quantum vacuum excitations. Depending on the material polytype, as is the case for SiC, plasmon polaritons with {\it longitudinal} or {\it hybrid longitudinal-transverse} polarizations are possible and they can be resolved simply by tracking the scaling law and/or sign of the Casimir interaction. From our study, we find that for separations less than 10 microns,the magnitude of Casimir pressure is around $10^{-8} - 10^{-3}$ mPa. This range is accessible experimentally, as demonstrated recently by the Casimir And Non-Newtonian force EXperiment (CANNEX) reporting measure Casimir pressures in the $10^{-8} - 10^{-3}$ mPa range  \cite{Almasi2015, Sedmik2021}. In fact, signatures of the interaction in terms of scaling law, magnitude, and/or sign can be used to probe the different plasmon polariton modes in piezoelectric materials and their structural differences. Our study further expands the materials perspective of Casimir phenomena by demonstrating the unique role of phonons and their coupling to electromagnetic excitations.

\begin{acknowledgments}
We acknowledge support from the US Department of Energy under Grant No. DE-FG02-06ER46297. P. R.-L. acknowledges support AYUDA PUENTE, URJC, and from Ministerio de Ciencia e Innovaci\'{o}n (Spain), Agencia Estatal de Investigaci\'{o}n, under project NAUTILUS (PID2022-139524NB-I00).
\end{acknowledgments}

\section*{Author Contributions}
L.M.W. conceived the idea and D.-N. Le performed the calculations. P. R.-L. and L.M.W. performed the analysis. D.-N. Le and L.M.W. wrote the manuscript. 

\section*{Competing Interests}
The authors declare no competing interests.

\section*{Code availability}
The paper does not report an original code or mathematical algorithm.

\newpage

\onecolumngrid

\setcounter{page}{1}
\setcounter{section}{0}
\setcounter{table}{0}
\setcounter{figure}{0}
\setcounter{equation}{0}

\renewcommand{\thepage}{S-\arabic{page}} 
\renewcommand{\thesection}{S-\Roman{section}}  
\renewcommand{\thetable}{S-\Roman{table}}  
\renewcommand{\thefigure}{S-\arabic{figure}}
\renewcommand{\theequation}{S-\arabic{equation}}

\begin{center}
\begin{large}
\textbf{
Supplementary Information:\\
Phonon-assisted Casimir interactions between piezoelectric materials
}
\end{large}
\end{center}

\section{\label{appA}Boundary conditions and Fresnel Reflection Matrices for Piezoelectric Plates with semi-infinite Thickness}

\begin{figure}[H]
\begin{center}
\includegraphics[width = 0.8 \columnwidth]{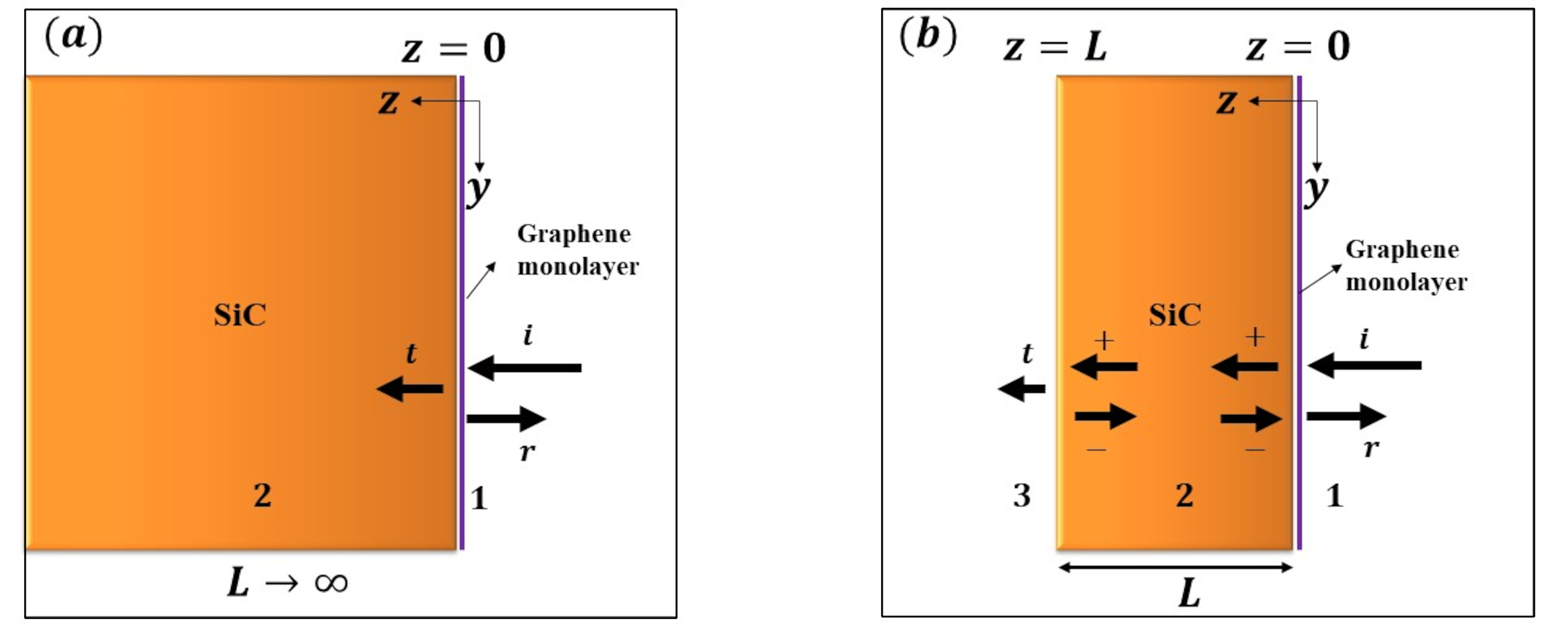}
\caption{\label{fig:S0} A diagram for the incoming, $i$, reflected, $r$, and transmistted, $t$ fields at the boundary of: (a) a semi-infinit SiC plate; (b) a SiC plate with a finite thickness $L$. Fields denoted with 
$+$ and $-$ correspond to propagations along incoming and reflected directions, respectively. The graphene monolayer is also shown.}
\end{center}
\end{figure}

To resolve the boundary conditions for the studied system, we consider the electromagnetic and ionic fields in medium 1 ($z < 0$) as shown in Fig.\ref{fig:S0}(a),
\begin{eqnarray}
    \label{Seqn:field-1}
    \left\{
\begin{array}{rcl}
\mathbf{D}_1 (\mathbf{r}, t) = \varepsilon_0 \mathbf{E}_1  (\mathbf{r}, t) & = & \varepsilon_0 \left\{ \left[ {\mathcal{E}}_s^i \pmb{\hat{x}} + {\mathcal{E}}_p^i \left( - \frac{q_z c}{ \omega} \pmb{\hat{y}} + \frac{q_{\parallel} c}{ \omega} \pmb{\hat{z}} \right) \right] e^{i q_z z} +  \left[ {\mathcal{E}}_s^r \pmb{\hat{x}} + {\mathcal{E}}_p^r \left( \frac{q_z c}{ \omega} \pmb{\hat{y}} + \frac{q_{\parallel} c}{ \omega} \pmb{\hat{z}} \right) \right] e^{-i q_z z} \right\} e^{i \left( q_{\parallel} y - \omega t\right)}, \\
c \mathbf{B}_1 (\mathbf{r},t) = c \mu_0 \mathbf{H}_1 (\mathbf{r},t) & = & \left\{ \left[{\mathcal{E}}_p^i \pmb{\hat{x}} - {\mathcal{E}}_s^i \left( - \frac{q_z c}{ \omega} \pmb{\hat{y}} + \frac{q_{\parallel} c}{ \omega} \pmb{\hat{z}} \right) \right] e^{i q_z z} + \left[{\mathcal{E}}_p^r \pmb{\hat{x}} - {\mathcal{E}}_s^r \left( \frac{q_z c}{ \omega} \pmb{\hat{y}} + \frac{q_{\parallel} c}{ \omega} \pmb{\hat{z}} \right) \right]  e^{-i q_z z} \right\} e^{i \left( q_{\parallel} y - \omega t\right)},\\
\mathbf{u}_1 (\mathbf{r}, t ) & = & \pmb{0},
\end{array}
    \right.
\end{eqnarray}
where $\mathbf{q}_{\parallel}=(q_x,q_y)$ and $q_{z} = \sqrt{\omega^2/c^2 - q_{\parallel}^2}$ are in-plane and out-of-plane components of the wave vector, respectively. The amplitudes ${\mathcal{E}}_s^i, {\mathcal{E}}_p^i$ correspond to the s- and p-polarized incoming electromagnetic waves, while ${\mathcal{E}}_s^r, {\mathcal{E}}_p^r$ are the amplitudes of the s- and p-polarized reflected electromagnetic waves. 

Inside the piezoelectric medium 2 $(z > 0)$ (Fig. \ref{fig:S0}(a)), the electromagnetic and ionic displacement fields are
\begin{eqnarray}
    \label{Seqn:field-2}
        \left\{
\begin{array}{rcl}
 \mathbf{E}_2(\mathbf{r}, t)  & = & \left\{ \left[ {\mathcal{E}}_{s,PhP} \pmb{\hat{x}} + {\mathcal{E}}_{p,PhP} \left( - \frac{q_{PhP,z} c}{ \omega} \pmb{\hat{y}} + \frac{q_{\parallel} c}{ \omega} \pmb{\hat{z}} \right) \right] e^{i q_{PhP,z} z} \right. \\
 &&\quad \quad \quad \left. - \mathcal{E}_{LO} \left( \frac{q_{\parallel} \beta_{LO}}{\omega} \pmb{\hat{y}} + \frac{q_{LO,z} \beta_{LO}}{\omega} \pmb{\hat{z}} \right)  e^{i q_{LO,z} z} \right\} e^{i \left( q_{\parallel} y - \omega t\right)}, \\
 \mathbf{D}_2(\mathbf{r}, t)  & = & \varepsilon_0 \varepsilon_{PhP} (\omega, q)  \left[ {\mathcal{E}}_{s,PhP} \pmb{\hat{x}} + {\mathcal{E}}_{p,PhP} \left( - \frac{q_{PhP,z} c}{ \omega} \pmb{\hat{y}} + \frac{q_{\parallel} c}{ \omega} \pmb{\hat{z}} \right) \right] e^{i q_{PhP,z} z} e^{i \left( q_{\parallel} y - \omega t\right)}, \\
 c \mathbf{B}_2(\mathbf{r}, t) = c \mu_0 \mathbf{H}_2(\mathbf{r}, t) & = & \left[ \varepsilon_{PhP} (\omega, q) {\mathcal{E}}_{p,PhP} \pmb{\hat{x}} - {\mathcal{E}}_{s,PhP} \left( - \frac{q_{PhP,z} c}{ \omega} \pmb{\hat{y}} + \frac{q_{\parallel} c}{ \omega} \pmb{\hat{z}} \right) \right] e^{i q_{PhP,z} z} e^{i \left( q_{\parallel} y - \omega t\right)} \\
\mathbf{u}_2(\mathbf{r}, t) & = & \varepsilon_0^{-1} e_i \left\{ \left[ {\mathcal{E}}_{s,TO} \pmb{\hat{x}} + {\mathcal{E}}_{p,TO} \left( - \frac{q_{TO,z} \beta_{TO}}{ \omega} \pmb{\hat{y}} + \frac{q_{\parallel} \beta_{TO}}{ \omega} \pmb{\hat{z}} \right) \right] e^{i q_{TO,z} z} \right. \\
&& \quad \quad \quad \left. + \mathcal{E}_{LO} \left( \frac{q_{\parallel} \beta_{LO}}{\omega} \pmb{\hat{y}} + \frac{q_{LO,z} \beta_{LO}}{\omega} \pmb{\hat{z}} \right) e^{i q_{LO,z} z} \right. \\
&& \quad \quad \quad \left. + \left[ {\mathcal{E}}_{s,PhP} \pmb{\hat{x}} + {\mathcal{E}}_{p,PhP} \left( - \frac{q_{PhP,z} c}{ \omega} \pmb{\hat{y}} + \frac{q_{\parallel} c}{ \omega} \pmb{\hat{z}} \right) \right] \left[ \varepsilon_{PhP} (\omega, q) - \varepsilon_{\infty} \right] e^{i q_{PhP,z} z} \right\} e^{i \left( q_{\parallel} y - \omega t\right)} 
\end{array}
    \right.
\end{eqnarray}
where $\mathcal{E}_{s,PhP}, \mathcal{E}_{p,PhP}$ are the amplitudes of the s- and p-polarized electromagnetic waves associated with the phonon polaritons inside the piezoelectric material and $\mathcal{E}_{LO}$ is the amplitude associated with the electric field induced by ionic vibration of the LO phonons. The amplitudes $\mathcal{E}_{s,TO}, \mathcal{E}_{p,TO}$ correspond to the s- and p-polarized electromagnetic waves associated with the TO phonon vibrations. Also, $q_{TO,z} = \sqrt{\omega_{TO}^2 - \omega^2}/\beta_{TO}$ and $q_{LO,z} = \sqrt{\omega_{LO}^2 - \omega^2}/\beta_{TO}$ are out-of-plane components of TO and LO phonon wave vectors.   

The coupled electromagnetic-ionic fields must satisfy two sets of boundary conditions simulteneously. One set are the standard Maxwell boundary conditions for the electromagnetic fields,
\begin{equation}
    \label{Seqn:MBC}
    \left\{
\begin{array}{lll}
 \mathbf{D}_{1,z} + \omega^{-1} q_{\parallel} \sigma_s \mathbf{E}_{1,y} & = &  \mathbf{D}_{2,z} \\
 \mathbf{E}_{1,x}  & = & \mathbf{E}_{2,x}, \\
 \mathbf{E}_{1,y}  & = & \mathbf{E}_{2,y}, \\
 \mathbf{B}_{1,z}  & = & \mathbf{B}_{2,z}, \\
 \mathbf{H}_{1,y} - \sigma_s \mathbf{E}_{1,x}  & = & \mathbf{H}_{2,y}, \\
 \mathbf{H}_{1,x} + \sigma_s \mathbf{E}_{1,y}  & = & \mathbf{H}_{2,x} .
\end{array}
    \right..
\end{equation}
One notes that the continuity conditions for $\mathbf{E}_x$ and $\mathbf{H}_x$ are equivalent to the ones for $\mathbf{B}_z$ and $\mathbf{D}_z$, respectively. The other set are the elastic boundary condition for the continuity of the ionic displacement fields across the boundary \cite{Ridley1994, Gubbin2020}
\begin{equation}
    \label{Seqn:EBC}
    \left\{
\begin{array}{lll}
 \mathbf{u}_{1,x} & = &  \mathbf{u}_{2,x} \\
 \mathbf{u}_{1,y} & = &  \mathbf{u}_{2,y} \\
 \mathbf{u}_{1,z} & = &  \mathbf{u}_{2,z} \\
\end{array}
    \right..
\end{equation}

The Fresnel reflection coefficients relate the incoming and reflected waves at the boundary, such that
\begin{eqnarray}
    \label{Seqn:fresnel}
    \begin{pmatrix}
        {\mathcal{E}}_s^r \\
        {\mathcal{E}}_p^r
    \end{pmatrix} = \begin{pmatrix}
        r_{ss} & r_{sp} \\
        r_{ps} & r_{pp} 
    \end{pmatrix} \begin{pmatrix}
        {\mathcal{E}}_s^i \\
        {\mathcal{E}}_p^i
    \end{pmatrix} .
\end{eqnarray}
 From the solution of Eqs. (S1-S4), we then find $r_{sp} = r_{ps} = 0$ while
\begin{eqnarray*}
    \label{Seqn:r-ss-final}
   \lim\limits_{L\to\infty} r_{ss} \left( \mathbf{q}_{\parallel},\omega \right) &=& \frac{q_z - q_{PhP,z} - \mu_0 \omega \sigma_s (\mathbf{q}_{\parallel},\omega)}{q_z + q_{PhP,z} + \mu_0 \omega \sigma_s (\mathbf{q}_{\parallel},\omega)} \\
    \label{Seqn:r-pp-final}
\lim\limits_{L\to\infty} r_{pp} \left( \mathbf{q}_{\parallel},\omega \right) &=& \frac{\varepsilon_{PhP}( \omega)q_z - \left( q_{PhP,z} + \Omega(\mathbf{q}_{\parallel},\omega) \right) + \mu_0 c \sigma_s (\mathbf{q}_{\parallel},\omega) \left(\frac{q_z  c}{\omega}\right) \left( q_{PhP,z} + \Omega(\mathbf{q}_{\parallel},\omega) \right) }{ \varepsilon_{PhP}( \omega)q_z + \left( q_{PhP,z} + \Omega(\mathbf{q}_{\parallel},\omega) \right) + \mu_0 c \sigma_s (\mathbf{q}_{\parallel},\omega) \left(\frac{q_z  c}{\omega}\right) \left( q_{PhP,z} + \Omega(\mathbf{q}_{\parallel},\omega) \right) },
\end{eqnarray*}
where 
\begin{eqnarray}
    \label{Seqn:E-LO}
    \Omega (\mathbf{q}_{\parallel}, \omega) = \frac{c\mathcal{E}_{p,PhP}}{q_{\parallel} \beta_{LO} \mathcal{E}_{LO}} = \frac{q_{\parallel}^2 \left( q_{PhP,z} - q_{TO,z} \right)}{ q_{\parallel}^2+ q_{TO,z} q_{LO,z} } \left( \varepsilon_{PhP}(\omega)  - \varepsilon_{\infty} \right),
\end{eqnarray}
as shown in Eq. (5) in the manuscript.

Let us also note, that one might consider another set of elastic boundary conditions, which correspond to the continuity of normal component of mechanical stress tensor \cite{Gubbin2019, Gubbin2020}
\begin{equation}
    \label{Seqn:EBC-another}
    \left\{
\begin{array}{lll}
\partial_z \mathbf{u}_{1,x} & = & \partial_z \mathbf{u}_{2,x} \\
\partial_z \mathbf{u}_{1,y} & = & \partial_z  \mathbf{u}_{2,y} \\
 \mathbf{u}_{1,z} & = &  \mathbf{u}_{2,z} \\
\end{array}
    \right..
\end{equation}
from which we find an alternative to Eq. \eqref{Seqn:EBC-another} solution,
\begin{equation}
\label{Seqn:Omega-LO-another}
\Omega^{\prime}(\mathbf{q}_{\parallel}, \omega) = \frac{q_{\parallel}^2 \left( q_{PhP,z}^2 - q_{TO,z}^2 \right)}{ \left(q_{\parallel}^2+ q_{TO,z}^2 \right) q_{LO,z} } \left( \varepsilon_{PhP}(\omega)  - \varepsilon_{\infty} \right) .
\end{equation}
Our calculations show that the difference in the Casimir interactions computed with $\Omega(\mathbf{q}_{\parallel}, \omega)$ and $\Omega^{\prime}(\mathbf{q}_{\parallel}, \omega)$ are minute. This is not surprising since at the long wavelength limit $q_{\parallel} \to 0$, the most important regime for Casimir phenomena, the two factors are indistinguishable,
\begin{equation}
\label{Seqn:Omega-LO-compare}
\lim\limits_{\omega \to \infty} \Omega(\mathbf{q}_{\parallel}, \omega) = \lim\limits_{\omega \to \infty} \Omega^{\prime}(\mathbf{q}_{\parallel}, \omega) = 0, \quad \lim\limits_{\omega \to 0} \Omega(\mathbf{q}_{\parallel}, \omega) = \lim\limits_{\omega \to 0} \Omega^{\prime}(\mathbf{q}_{\parallel}, \omega) = - \frac{ q_{\parallel}^2 (\varepsilon_{st} - \varepsilon_{\infty}) }{q_{LO,z}}.
\end{equation}
Thus, in all calculations, we show results associated with the elastic boundary conditions with  $\Omega(\mathbf{q}_{\parallel}, \omega)$  in \eqref{Seqn:EBC}.

\subsection{Zone folding of LO phonons}
Additional Bragg peaks in the spectra of some SiC polytypes is described by replacing the wave vector of the typical LO phonon mode $q_{\parallel}$ by $q_{\parallel}^{\prime} = q_{\parallel} + q_{n,m}$, where $q_{n,m}=\frac{2m\pi}{nc}$ ($m=1,2,...,n-1$ - number of Brag peak, $c$-elongated lattice constant of SiC polytype). Due to the propagation of the zone folded LO phonons, there is a phase factor $e^{i q_{LO,z} z} e^{i (q_{\parallel}^{\prime} y - \omega t)}$. We note that $e^{i q_{n,m}y}$ does not affect the ionic displacement field but it causes a mismatch in the electromagnetic fields. Following the arguments in  Ref. \cite{Gubbin2019}, this mismatch factor can be found by macroscopic averaging over the unit cell (UC) of lattice, such that
\begin{eqnarray}
    \label{eqn:average}
    \left< e^{i q_{n,m} y} \right> = \frac{1}{V_{c}} \iint\limits_{UC} e^{i q_{n,m} y} dx dy = \frac{2i}{\pi} \equiv \gamma.
\end{eqnarray}
The electromagnetic and ionic fields then become,
\begin{eqnarray}
    \label{Seqn:field-2-ZFLO}
        \left\{
\begin{array}{rcl}
 \mathbf{E}_2(\mathbf{r}, t)  & = & \left\{ \left[ {\mathcal{E}}_{s,PhP} \pmb{\hat{x}} + {\mathcal{E}}_{p,PhP} \left( - \frac{q_{PhP,z} c}{ \omega} \pmb{\hat{y}} + \frac{q_{\parallel} c}{ \omega} \pmb{\hat{z}} \right) \right] e^{i q_{PhP,z} z} \right. \\
&& \quad \quad \quad \left. - \mathcal{E}_{LO} \left( \frac{q_{\parallel}^{\prime} \beta_{LO}}{\omega} \pmb{\hat{y}} + \frac{q_{LO,z} \beta_{LO}}{\omega} \pmb{\hat{z}} \right)  e^{i q_{LO,z} z} \right\} e^{i \left( q_{\parallel} y - \omega t\right)}, \\
 \mathbf{D}_2(\mathbf{r}, t)  & = & \varepsilon_0 \varepsilon_{PhP} (\omega, q)  \left[ {\mathcal{E}}_{s,PhP} \pmb{\hat{x}} + {\mathcal{E}}_{p,PhP} \left( - \frac{q_{PhP,z} c}{ \omega} \pmb{\hat{y}} + \frac{q_{\parallel} c}{ \omega} \pmb{\hat{z}} \right) \right] e^{i q_{PhP,z} z} e^{i \left( q_{\parallel} y - \omega t\right)}, \\
 c \mathbf{B}_2(\mathbf{r}, t) = c \mu_0 \mathbf{H}_2(\mathbf{r}, t) & = & \left[ \varepsilon_{PhP} (\omega, q) {\mathcal{E}}_{p,PhP} \pmb{\hat{x}} - {\mathcal{E}}_{s,PhP} \left( - \frac{q_{PhP,z} c}{ \omega} \pmb{\hat{y}} + \frac{q_{\parallel} c}{ \omega} \pmb{\hat{z}} \right) \right] e^{i q_{PhP,z} z} e^{i \left( q_{\parallel} y - \omega t\right)} \\
\mathbf{u}_2(\mathbf{r}, t) & = & \varepsilon_0^{-1} e_i \left\{ \left[ {\mathcal{E}}_{s,TO} \pmb{\hat{x}} + {\mathcal{E}}_{p,TO} \left( - \frac{q_{TO,z} \beta_{TO}}{ \omega} \pmb{\hat{y}} + \frac{q_{\parallel} \beta_{TO}}{ \omega} \pmb{\hat{z}} \right) \right] e^{i q_{TO,z} z} \right.\\
&& \left. + \gamma^{-1} \mathcal{E}_{LO} \left( \frac{q_{\parallel}^{\prime} \beta_{LO}}{\omega} \pmb{\hat{y}} + \frac{q_{LO,z} \beta_{LO}}{\omega} \pmb{\hat{z}} \right) e^{i [q_{LO,z} z + (q_{\parallel}^{\prime}-q_{\parallel})y]} \right. \\
&& \quad \quad \quad \left. + \left[ {\mathcal{E}}_{s,PhP} \pmb{\hat{x}} + {\mathcal{E}}_{p,PhP} \left( - \frac{q_{PhP,z} c}{ \omega} \pmb{\hat{y}} + \frac{q_{\parallel} c}{ \omega} \pmb{\hat{z}} \right) \right] \left[ \varepsilon_{PhP} (\omega, q) - \varepsilon_{\infty} \right] e^{i q_{PhP,z} z} \right\} e^{i \left( q_{\parallel} y - \omega t\right)} 
\end{array}
    \right..
\end{eqnarray}
with a modified \eqref{Seqn:EBC} 
\begin{equation}
\label{Seqn:Omega-ZFLO}
\Omega_{ZFLO} (\mathbf{q}_{\parallel}, \omega) = \frac{\gamma q_{\parallel} q_{\parallel}^{\prime} \left( q_{PhP,z} - q_{TO,z} \right)}{ q_{\parallel} q_{\parallel}^{\prime}+ q_{TO,z} q_{LO,z} } \left( \varepsilon_{PhP}(\omega)  - \varepsilon_{\infty} \right) .
\end{equation}

Note that above formula captures both. Taking $q_{\parallel}^{\prime} = q_{\parallel}$, $\gamma = 1$ corresponds to the SPhPs, while taking $q_{\parallel}^{\prime} = q_{\parallel} + q_{m,n}$ and $\gamma = 2i/\pi$ corresponds to the hybrid LT-SPhPs.

\section{\label{appB}Boundary conditions and Fresnel Reflection Coefficients for Piezoelectric Plates with Finite Thickness}

To obtain the boundary conditions for piezoelectric plates with finite thickness (see Fig.\ref{fig:S0}(b)), we note that the electromagnetic and ionic displacement fields in medium 1 ($z < 0$) remains the same as Eq. \eqref{Seqn:field-1}, while the transmitted electromagnetic and ionic displacement fields at medium 3 $(z > L)$ are 
\begin{eqnarray}
    \label{Seqn:field-3}
    \left\{
\begin{array}{rcl}
\mathbf{D}_3 (\mathbf{r}, t) = \varepsilon_0 \mathbf{E}_3  (\mathbf{r}, t) & = & \varepsilon_0  \left[ {\mathcal{E}}_s^t \pmb{\hat{x}} + {\mathcal{E}}_p^t \left( - \frac{q_z c}{ \omega} \pmb{\hat{y}} + \frac{q_{\parallel} c}{ \omega} \pmb{\hat{z}} \right) \right] e^{i q_z z}  e^{i \left( q_{\parallel} y - \omega t\right)}  \\
c \mathbf{B}_3 (\mathbf{r},t) = c \mu_0 \mathbf{H}_3 (\mathbf{r},t) & = & \left[{\mathcal{E}}_p^t \pmb{\hat{x}} - {\mathcal{E}}_s^t \left( - \frac{q_z c}{ \omega} \pmb{\hat{y}} + \frac{q_{\parallel} c}{ \omega} \pmb{\hat{z}} \right) \right] e^{i q_z z} e^{i \left( q_{\parallel} y - \omega t\right)}\\
\mathbf{u}_3 (\mathbf{r}, t ) & = & \pmb{0}
\end{array}
    \right.
\end{eqnarray}
where ${\mathcal{E}}_s^t, {\mathcal{E}}_p^t$ are amplitudes of the s- and p-polarized transmitted electromagnetic waves.

Due to the multi-reflections inside the piezoelectric plate, the electromagnetic and ionic displacement fields inside the plate $0 < z < L$ is now the superposition of waves along $\pm z$ directions
\begin{eqnarray}
    \label{Seqn:field-2-finite}
        \left\{
\begin{array}{rcl}
 \mathbf{E}_2(\mathbf{r}, t)  & = & \left\{ \left[ {\mathcal{E}}_{s,PhP}^{+} \pmb{\hat{x}} + {\mathcal{E}}_{p,PhP}^{+} \left( - \frac{q_{PhP,z} c}{ \omega} \pmb{\hat{y}} + \frac{q_{\parallel} c}{ \omega} \pmb{\hat{z}} \right) \right] e^{i q_{PhP,z} z} \right. \nonumber\\
 && \quad \left. - \mathcal{E}_{LO}^{+} \left( \frac{q_{\parallel}^{\prime} \beta_{LO}}{\omega} \pmb{\hat{y}} + \frac{q_{LO,z} \beta_{LO}}{\omega} \pmb{\hat{z}} \right)  e^{i q_{LO,z} z} \right. \\ 
 && \quad \left. + \left[ {\mathcal{E}}_{s,PhP}^{-} \pmb{\hat{x}} + {\mathcal{E}}_{p,PhP}^{-} \left( \frac{q_{PhP,z} c}{ \omega} \pmb{\hat{y}} + \frac{q_{\parallel} c}{ \omega} \pmb{\hat{z}} \right) \right] e^{-i q_{PhP,z} z} \right. \nonumber\\
 && \quad \left. - \mathcal{E}_{LO}^{-} \left(\frac{q_{\parallel}^{\prime} \beta_{LO}}{\omega} \pmb{\hat{y}} - \frac{q_{LO,z} \beta_{LO}}{\omega} \pmb{\hat{z}} \right)  e^{-i q_{LO,z} z}\right\} e^{i \left( q_{\parallel} y - \omega t\right)}, \\
 \mathbf{D}_2(\mathbf{r}, t)  & = & \varepsilon_0 \varepsilon_{PhP} (\omega, q) \left\{ \left[ {\mathcal{E}}_{s,PhP}^{+} \pmb{\hat{x}} + {\mathcal{E}}_{p,PhP}^{+} \left( - \frac{q_{PhP,z} c}{ \omega} \pmb{\hat{y}} + \frac{q_{\parallel} c}{ \omega} \pmb{\hat{z}} \right) \right] e^{i q_{PhP,z} z} \right. \\
 && \quad \quad \quad \quad \quad \quad \left. + \left[ {\mathcal{E}}_{s,PhP}^{-} \pmb{\hat{x}} + {\mathcal{E}}_{p,PhP}^{-} \left(  \frac{q_{PhP,z} c}{ \omega} \pmb{\hat{y}} + \frac{q_{\parallel} c}{ \omega} \pmb{\hat{z}} \right) \right] e^{-i q_{PhP,z} z} \right\} e^{i \left( q_{\parallel} y - \omega t\right)}, \\
 c \mathbf{B}_2(\mathbf{r}, t) = c \mu_0 \mathbf{H}_2(\mathbf{r}, t) & = & \left\{ \left[ \varepsilon_{PhP} (\omega, q) {\mathcal{E}}_{p,PhP}^{+} \pmb{\hat{x}} - {\mathcal{E}}_{s,PhP}^{+} \left( - \frac{q_{PhP,z} c}{ \omega} \pmb{\hat{y}} + \frac{q_{\parallel} c}{ \omega} \pmb{\hat{z}} \right) \right] e^{i q_{PhP,z} z} \right. \\
&&  \left. + \left[ \varepsilon_{PhP} (\omega, q) {\mathcal{E}}_{p,PhP}^{-} \pmb{\hat{x}} - {\mathcal{E}}_{s,PhP}^{-} \left( \frac{q_{PhP,z} c}{ \omega} \pmb{\hat{y}} + \frac{q_{\parallel} c}{ \omega} \pmb{\hat{z}} \right) \right] e^{-i q_{PhP,z} z} \right\} e^{i \left( q_{\parallel} y - \omega t\right)} \\
\mathbf{u}_2(\mathbf{r}, t) & = & \varepsilon_0^{-1} e_i \left\{ \left[ {\mathcal{E}}_{s,TO}^{+} \pmb{\hat{x}} + {\mathcal{E}}_{p,TO}^{+} \left( - \frac{q_{TO,z} \beta_{TO}}{ \omega} \pmb{\hat{y}} + \frac{q_{\parallel} \beta_{TO}}{ \omega} \pmb{\hat{z}} \right) \right] e^{i q_{TO,z} z} \right. \\
&& \left. + \gamma^{-1} \mathcal{E}_{LO}^{+} \left( \frac{q_{\parallel}^{\prime} \beta_{LO}}{\omega} \pmb{\hat{y}} + \frac{q_{LO,z} \beta_{LO}}{\omega} \pmb{\hat{z}} \right) e^{i [q_{LO,z} z  (q_{\parallel}^{\prime} - q_{\parallel}) y]} \right. \\
&& \left. + \left[ {\mathcal{E}}_{s,PhP}^{+} \pmb{\hat{x}} + {\mathcal{E}}_{p,PhP}^{+} \left( - \frac{q_{PhP,z} c}{ \omega} \pmb{\hat{y}} + \frac{q_{\parallel} c}{ \omega} \pmb{\hat{z}} \right) \right] \left[ \varepsilon_{PhP} (\omega, q) - \varepsilon_{\infty} \right] e^{i q_{PhP,z} z} \right. \\
&& \left. + \left[ {\mathcal{E}}_{s,TO}^{-} \pmb{\hat{x}} + {\mathcal{E}}_{p,TO}^{-} \left(  \frac{q_{TO,z} \beta_{TO}}{ \omega} \pmb{\hat{y}} + \frac{q_{\parallel} \beta_{TO}}{ \omega} \pmb{\hat{z}} \right) \right] e^{-i q_{TO,z} z} \right.\\
&& \left. + \gamma^{-1} \mathcal{E}_{LO}^{-} \left( \frac{q_{\parallel}^{\prime} \beta_{LO}}{\omega} \pmb{\hat{y}} - \frac{q_{LO,z} \beta_{LO}}{\omega} \pmb{\hat{z}} \right)  e^{i [ - q_{LO,z} z  (q_{\parallel}^{\prime} - q_{\parallel}) y]} \right. \\
&& \left. + \left[ {\mathcal{E}}_{s,PhP}^{-} \pmb{\hat{x}} + {\mathcal{E}}_{p,PhP}^{-} \left( \frac{q_{PhP,z} c}{ \omega} \pmb{\hat{y}} + \frac{q_{\parallel} c}{ \omega} \pmb{\hat{z}} \right) \right] \left[ \varepsilon_{PhP} (\omega, q) - \varepsilon_{\infty} \right] e^{-i q_{PhP,z} z} 
\right\} e^{i \left( q_{\parallel} y - \omega t\right)} 
\end{array}
    \right.
\end{eqnarray}

The Maxwell boundary conditions at $z = 0$ (with surface conductivity $\sigma_s$) and $z = L$ (with no surface conductivity) interfaces are
\begin{equation}
    \label{Seqn:MBC-finite}
    \left\{
\begin{array}{lll}
 \mathbf{D}_{1,z} (z=0) + \omega^{-1} q_{\parallel} \sigma_s \mathbf{E}_{1,y} (z=0) & = &  \mathbf{D}_{2,z} (z=0) \\
 \mathbf{E}_{1,x} (z=0)  & = & \mathbf{E}_{2,x} (z=0), \\
 \mathbf{E}_{1,y} (z=0)  & = & \mathbf{E}_{2,y} (z=0), \\
 \mathbf{B}_{1,z} (z=0) & = & \mathbf{B}_{2,z} (z=0), \\
 \mathbf{H}_{1,y} (z=0) - \sigma_s \mathbf{E}_{1,x} (z=0) & = & \mathbf{H}_{2,y} (z=0), \\
 \mathbf{H}_{1,x} (z=0) + \sigma_s \mathbf{E}_{1,y}  & = & \mathbf{H}_{2,x} (z=0) \\
  \mathbf{D}_{3,z} (z=L) & = &  \mathbf{D}_{2,z} (z=L) \\
 \mathbf{E}_{3,x} (z=L)  & = & \mathbf{E}_{2,x} (z=L), \\
 \mathbf{E}_{3,y} (z=L)  & = & \mathbf{E}_{2,y} (z=L), \\
 \mathbf{B}_{3,z} (z=L) & = & \mathbf{B}_{2,z} (z=L), \\
 \mathbf{H}_{3,y} (z=L) & = & \mathbf{H}_{2,y} (z=L), \\
 \mathbf{H}_{3,x} (z=L)  & = & \mathbf{H}_{2,x} (z=L)
\end{array}
    \right..
\end{equation}
The elastic boundary conditions read
\begin{equation}
    \label{Seqn:EBC-finite}
    \left\{
\begin{array}{lll}
 \mathbf{u}_{1,x} (z = 0) & = &  \mathbf{u}_{2,x} (z = 0)\\
 \mathbf{u}_{1,y} (z = 0) & = &  \mathbf{u}_{2,y} (z = 0) \\
 \mathbf{u}_{1,z} (z = 0) & = &  \mathbf{u}_{2,z} (z = 0) \\
  \mathbf{u}_{3,x} (z = 0) & = &  \mathbf{u}_{2,x} (z = L)\\
 \mathbf{u}_{3,y} (z = 0) & = &  \mathbf{u}_{2,y} (z = L) \\
 \mathbf{u}_{3,z} (z = 0) & = &  \mathbf{u}_{2,z} (z = L) 
\end{array}
    \right..
\end{equation}

Solving the elastic boundary condition \eqref{Seqn:EBC-finite}, we obtained the following relations
\begin{eqnarray}
    \label{eqn:E-LO-finite-relation}
   q_{\parallel}^{\prime} \beta_{LO} \mathcal{E}_{LO}^{\pm} = \Omega_{\pm,+} (\mathbf{q}_{\parallel}, \omega) c\mathcal{E}_{p,PhP}^{+} + \Omega_{\pm,-} (\mathbf{q}_{\parallel}, \omega) c\mathcal{E}_{p,PhP}^{-}.
\end{eqnarray}

Consequently, we find tha the Fresnel reflection coefficients at each boundary for the $s-$polarized modes are,

\begin{eqnarray}
    \label{Seqn:rss-12-finite}
    r_{ss}^{12} &=& \frac{q_z - q_{PhP,z} - \mu_0 \omega \sigma_s }{q_z + q_{PhP,z} + \mu_0 \omega \sigma_s },\\
    \label{Seqn:rss-21-finite}
    r_{ss}^{21}  &=& - \frac{q_z - q_{PhP,z} + \mu_0 \omega \sigma_s}{q_z + q_{PhP,z} + \mu_0 \omega \sigma_s}, \\
    \label{Seqn:rss-21-23-finite}
    r_{ss}^{23}  &=& - \frac{q_z - q_{PhP,z} }{q_z + q_{PhP,z} }.
\end{eqnarray}

The Fresnel reflection coefficients at each boundary for the $p-$polarized modes are,

\begin{eqnarray}
    r_{pp}^{12}  &=& \frac{\varepsilon_{PhP}q_z - \left( q_{PhP,z} + \Omega_{+,+} + \Omega_{-,+}  \right)  \left( 1 - \mu_0 c \sigma_s   \left(\frac{q_z   c}{\omega}\right) \right) }{ \varepsilon_{PhP}q_z + \left( q_{PhP,z} + \Omega_{+,+}  + \Omega_{-,+}  \right)  \left( 1 + \mu_0 c \sigma_s   \left(\frac{q_z   c}{\omega}\right) \right) }, \\
    t_{pp}^{12}  &=& \frac{2 q_z}{ \varepsilon_{PhP} q_z + \left( q_{PhP,z} + \Omega_{+,+}  + \Omega_{-,+}  \right)  \left( 1 + \mu_0 c \sigma_s   \left(\frac{q_z   c}{\omega}\right) \right) }, \\
    r_{pp}^{21}  &=& - \frac{\varepsilon_{PhP} q_z - \left( q_{PhP,z} - \Omega_{+,-}  - \Omega_{-,-}  \right)  \left( 1 + \mu_0 c \sigma_s   \left(\frac{q_z   c}{\omega}\right) \right) }{ \varepsilon_{PhP} q_z + \left( q_{PhP,z} + \Omega_{+,+}  + \Omega_{-,+}  \right)  \left( 1 + \mu_0 c \sigma_s   \left(\frac{q_z   c}{\omega}\right) \right) }, \\
    t_{pp}^{21}  &=&  \frac{ \varepsilon_{PhP}\left[ \left( q_{PhP,z} - \Omega_{+,-}  - \Omega_{-,-}  \right) + \left( q_{PhP,z} + \Omega_{+,+}  + \Omega_{-,+}  \right) \right] }{ \varepsilon_{PhP} q_z + \left( q_{PhP,z} + \Omega_{+,+}  + \Omega_{-,+}  \right)  \left( 1 + \mu_0 c \sigma_s   \left(\frac{q_z   c}{\omega}\right) \right) }, \\
    r_{pp}^{23}  &=& - \frac{\varepsilon_{PhP}q_z - \left( q_{PhP,z} + \Omega_{+,+}  e^{i(q_{LO,z} - q_{PhP,z}) L} + \Omega_{-,+}  e^{-i(q_{LO,z} + q_{PhP,z}) L} \right)   }{ \varepsilon_{PhP}q_z + \left( q_{PhP,z} - \Omega_{+,-}  e^{i(q_{LO,z} + q_{PhP,z}) L} - \Omega_{-,-}  e^{-i(q_{LO,z} - q_{PhP,z}) L} \right)  },\\
    t_{pp}^{23}  &=& \frac{\varepsilon_{PhP} \left[ 2 q_{PhP,z}  + \Omega_{+,+}  e^{i(q_{LO,z} - q_{PhP,z}) L} + \Omega_{-,+}  e^{-i(q_{LO,z} + q_{PhP,z}) L} - \Omega_{+,-}  e^{i(q_{LO,z} + q_{PhP,z}) L} - \Omega_{-,-}  e^{-i(q_{LO,z} - q_{PhP,z}) L} \right] }{ \varepsilon_{PhP}q_z  + \left( q_{PhP,z}  - \Omega_{+,-}  e^{i(q_{LO,z} + q_{PhP,z}) L} - \Omega_{-,-}  e^{-i(q_{LO,z} - q_{PhP,z}) L} \right) }, \nonumber\\
\end{eqnarray}
where $\Omega_{\pm,\pm}$ are obtained as
\begin{eqnarray}
    \label{Seqn:Omega-plus-finite}
    \Omega_{+,\pm} &=& \gamma q_{\parallel} q_{\parallel}^{\prime} \left[- 2 q_{TO,z} \left( - q_{\parallel} q_{\parallel}^{\prime} \pm q_{PhP,z} q_{LO,z} \right) \left( 1 + e^{i(-q_{LO,z} \pm q_{PhP,z})L} \right)  \right. \nonumber\\
    && \quad \quad \quad + \left(\pm q_{PhP,z} - q_{TO,z} \right) \left( q_{LO,z} q_{TO,z} + q_{\parallel} q_{\parallel}^{\prime} \right) \left( e^{-i(q_{LO,z} + q_{TO,z}) L} + e^{i(q_{TO,z} \pm q_{PhP,z})L} \right) \nonumber \\
    && \quad \quad \quad + \left. \left(\pm q_{PhP,z} + q_{TO,z} \right) \left( q_{LO,z} q_{TO,z} - q_{\parallel} q_{\parallel}^{\prime} \right) \left( e^{-i(q_{LO,z} - q_{TO,z}) L} + e^{i(-q_{TO,z} \pm q_{PhP,z})L} \right) \right]  \left( \varepsilon_{PhP} (\omega) - \varepsilon_{\infty} \right) \nonumber\\
    && \times \left[- 4 \left( q_{TO,z}^2 q_{LO,z}^2 + q_{\parallel}^2 (q_{\parallel}^{\prime})^2 \right) \sin ( q_{TO,z} L ) \sin ( q_{LO,z} L ) + 8 q_{LO,z} q_{TO,z} q_{\parallel} q_{\parallel}^{\prime} \left( \cos ( q_{TO,z} L ) \cos ( q_{LO,z} L )  - 1 \right) \right]^{-1} \nonumber\\
    \\
    \label{Seqn:Omega-minus-finite}
    \Omega_{-,\pm} &=& \gamma q_{\parallel} q_{\parallel}^{\prime} \left[ - 2 q_{TO,z} \left( q_{\parallel} q_{\parallel}^{\prime} \pm q_{PhP,z} q_{LO,z} \right) \left( 1 + e^{i(q_{LO,z} \pm q_{PhP,z})L} \right) \right. \nonumber\\
    && \quad \quad \quad  + \left(\pm q_{PhP,z} + q_{TO,z} \right) \left( q_{LO,z} q_{TO,z} + q_{\parallel} q_{\parallel}^{\prime} \right) \left( e^{i(q_{LO,z} + q_{TO,z}) L} + e^{i(-q_{TO,z} \pm q_{PhP,z})L} \right) \nonumber\\ 
    && \quad \quad \quad + \left. \left(\pm q_{PhP,z} - q_{TO,z} \right) \left( q_{LO,z} q_{TO,z} - q_{\parallel} q_{\parallel}^{\prime} \right) \left( e^{i(q_{LO,z} - q_{TO,z}) L} + e^{i(q_{TO,z} \pm q_{PhP,z})L} \right) \right]  \left( \varepsilon_{PhP} (\omega) - \varepsilon_{\infty} \right) \nonumber\\
    && \times \left[- 4 \left( q_{TO,z}^2 q_{LO,z}^2 + q_{\parallel}^2 (q_{\parallel}^{\prime})^2 \right) \sin ( q_{TO,z} L ) \sin ( q_{LO,z} L ) + 8 q_{LO,z} q_{TO,z} q_{\parallel} q_{\parallel}^{\prime} \left( \cos ( q_{TO,z} L ) \cos ( q_{LO,z} L )  - 1 \right) \right]^{-1} .\nonumber\\
\end{eqnarray}
Here $q_{\parallel}^{\prime} = q_{\parallel}$ and $\gamma = 1$  for the SPhP modes while $q_{\parallel}^{\prime} = q_{\parallel} + q_{m,n}$ and $\gamma = 2i/\pi$ for LT-SPhP modes;  $q_{PhP,z}$, $q_{LO,z}$, $q_{TO,z}$ are out-of-plane wave vector of PhP, LO and TO modes.

\section{\label{appC}Fresnel reflection coefficients of semi-infinite plates at the long-wavelength limit}

In the limit of $\omega \to 0$, the $s$- and $p$-polarized Fresnel reflection coefficients given in Eqs. (3)--(4) without graphene, we find
\begin{eqnarray}
    \label{Seqn:r-ss-0}
   \lim\limits_{L\to\infty} r_{ss} \left( \mathbf{q}_{\parallel},0 \right) &=& 0 \\
    \label{Seqn:r-pp-0}
\lim\limits_{L\to\infty} r_{pp} \left( \mathbf{q}_{\parallel},0 \right) &=& \frac{\varepsilon_{st} - \left(1 - i q_{\parallel}^{-1} \Omega(\mathbf{q}_{\parallel}, 0)\right) }{ \varepsilon_{st} - \left( 1 - i q_{\parallel}^{-1} \Omega(\mathbf{q}_{\parallel},0) \right) } = \dfrac{\dfrac{\varepsilon_{st}}{1 - i q_{\parallel}^{-1} \Omega(\mathbf{q}_{\parallel}, 0)} - 1 }{\dfrac{\varepsilon_{st}}{1 - i q_{\parallel}^{-1} \Omega(\mathbf{q}_{\parallel}, 0)} + 1 }.
\end{eqnarray}
As can be seen from the above expressions, the $p$-polarized Fresnel reflection coefficient corresponds to a medium with effective constant dielectric
\begin{equation}
\label{Seqn:effective-eps}
\varepsilon_{eff} = \dfrac{\varepsilon_{st}}{1 - i q_{\parallel}^{-1} \Omega(\mathbf{q}_{\parallel}, 0)} .
\end{equation}
For the piezoelectric semi-infinite plate, then, the contribution of the electromagnetic-elastic boundary conditions of LO phonons at long-wavelength limit $q_{\parallel}\to 0$ is $\Omega(\mathbf{q}_{\parallel}, 0) \propto q_{\parallel}^2$, thus, $q_{\parallel}^{-1} \Omega(\mathbf{q}_{\parallel}, 0) \propto q_{\parallel}$ while low frequency limit of the dielectric function is simply $\varepsilon_{eff} \rightarrow \varepsilon_{st}$.

For a semi-infinite plate with zone-folded LO phonons at the $q_{\parallel}\to 0$, we have 
\begin{equation}
\label{Seqn:effective-eps-ZFLO}
\lim\limits_{q \to 0}\Omega_{ZFLO}(\mathbf{q}_{\parallel}, 0) = - \frac{2i}{\pi} \frac{q_{\parallel} q_{1,2} \beta_{LO} }{\sqrt{\omega_{LO}^2 - \beta_{LO}^2 q_{1,2}^2}} \left( \varepsilon_{st} - \varepsilon_{\infty} \right) \Rightarrow \varepsilon_{eff} = \dfrac{\varepsilon_{st}}{1 - \frac{2}{\pi} \frac{q_{1,2} \beta_{LO} }{\sqrt{\omega_{LO}^2 - \beta_{LO}^2 q_{1,2}^2}} \left( \varepsilon_{st} - \varepsilon_{\infty} \right)} \approx 12.6 > \varepsilon_{st} = 10.
\end{equation}
Here $\omega_{TO} = 795 \text{ cm}^{-1}$, $\varepsilon_{\infty} = 6.5$, $\varepsilon_{st} = 10.0$, $\beta_{TO} = 9.15 \times 10^5 \text{ cm}/\text{s}$, $\beta_{LO} = 15.39 \times 10^5 \text{ cm}/\text{s}$, $\gamma = 4 \text{ cm}^{-1}$, $\omega_{LO} = \omega_{TO} \sqrt{\varepsilon_{st} / \varepsilon_{\infty}}$ and $q_{1,2} = \pi / c_2$ with $c_2 = 5 \text{ \AA}$ are taken for 4H-SiC \cite{Gubbin2019}.

In the case of graphene monolayers on top of the piezoelectric surfaces, for the low frequency regime $\omega \to 0$, the $s$- and $p$-polarized Fresnel reflection coefficients become 
\begin{eqnarray}
    \label{Seqn:r-ss-0-graphene}
   \lim\limits_{L\to\infty} r_{ss} \left( \mathbf{q}_{\parallel},0 \right) &=& 0 \\
    \label{Seqn:r-pp-0-graphene}
\lim\limits_{L\to\infty} r_{pp} \left( \mathbf{q}_{\parallel},0 \right) &=& 1.
\end{eqnarray}

\section{\label{appD}Fresnel reflection coefficients of Piezoelectric Plates with Finite Thickness at the long-wavelength limit}

Here we explicitly examine when there is no graphene cover.

\subsection{Finite Thickness Piezoelectric Plates assisted by SPhP modes}

At the long wavelength limit, the elastic boundary terms associated with SPhP modes are proportional to the square of in-plane wave vector since $\Omega_{\pm,\pm} \propto q_{\parallel}^2$. Therefore, these terms do not contribute at the $q \rightarrow 0$ limit. As the result, the following relations between $r^{ij}_{pp}$, $t^{ij}_{pp}$ are found,
\begin{equation}
\label{Seqn:rpp-relation-SPhP}
t_{pp}^{12} = \frac{r_{pp}^{12}+1}{\varepsilon_{PhP}}, \quad t_{pp}^{21} = \varepsilon_{PhP} \left( r_{pp}^{21}+1 \right), \quad r_{pp}^{21} = r_{pp}^{23} = - r_{pp}^{12},
\end{equation}  
and the p-polarized Fresnel reflection coefficient becomes
\begin{eqnarray}
\label{Seqn:rpp-SPhP-finite}
r_{pp} (\mathbf{q}_{\parallel},i \xi)= \frac{r_{pp}^{12} (\mathbf{q}_{\parallel},i \xi) \left(1 - e^{- 2 \kappa_{PhP,z}L} \right)}{1- (r_{pp}^{12} (\mathbf{q}_{\parallel},i \xi))^2 e^{-2 \kappa_{PhP,z}L}} \xrightarrow{L \to 0} 2 L\left( \frac{\varepsilon_{st}-1}{4 \varepsilon_{st}} \right) \dfrac{(\varepsilon_{st} + 1) \xi^2 / c^2 + \varepsilon_{st} {q}_{\parallel}^2  }{ \sqrt{ \xi^2 / c^2 + {q}_{\parallel}^2} }.
\end{eqnarray}
Similarly, we find that the s-polarized Fresnel reflection coefficient is
\begin{eqnarray}
\label{Seqn:rss-SPhP-finite}
r_{ss} (\mathbf{q}_{\parallel},i \xi)= \frac{r_{pp}^{12} (\mathbf{q}_{\parallel},i \xi) \left(1 - e^{- 2 \kappa_{PhP,z}L} \right)}{1- (r_{pp}^{12} (\mathbf{q}_{\parallel},i \xi))^2 e^{-2 \kappa_{PhP,z}L}} \xrightarrow{L \to 0} - 2 L\left( \frac{\varepsilon_{st}-1}{4} \right) \dfrac{\xi^2 / c^2  }{ \sqrt{ \xi^2 / c^2 + {q}_{\parallel}^2} }.
\end{eqnarray}

Consequently, the Casimir pressure at quantum limit i.e $T = 0$ is found with $D^{-6}$ scaling law
\begin{eqnarray}
    \label{Seqn:Casimir-quantum-SPhP-finite}
P_{qm} (D \gg L) = \frac{15 (\varepsilon_{st}-1)^2 \left( 9 \varepsilon_{st}^2 + 10 \varepsilon_{st} + 4 \right)L^2 }{2 \pi^4 \varepsilon_{st}^2 D^2} P_m = - \frac{\hbar c  (\varepsilon_{st}-1)^2 \left( 9 \varepsilon_{st}^2 + 10 \varepsilon_{st} + 4 \right) L^2}{32 \pi^2 \varepsilon_{st}^2 D^6}.
\end{eqnarray}
Also, the Casimir pressure at the thermal limit, which is only contributed from p-polarization, is found with $D^{-5}$ scaling law
\begin{eqnarray}
    \label{Seqn:Casimir-thermal-SPhP-finite}
P_{th} (D \gg L) =  - \frac{3 (\varepsilon_{st}-1)^2 L^2 k_B T}{32 \pi D^5}.
\end{eqnarray}
These scaling laws agree well with the numerical calculations given in Section \ref{appE}.

\subsection{Finite Thickness Piezoelectric Plates assisted by LT-SPhP modes}

Unlike the case of SPhP modes, the elastic boundary terms associated with LT-SPhPs are linearly proportional to in-plane wave vector i.e $\Omega_{\pm,\pm} \propto q_{\parallel}$. More explicitly we find
\begin{eqnarray}
\label{Seqn:Omega-plus-relation-LTSPhP}
\frac{\Omega_{+,\pm}}{i q_{\parallel}} & \xrightarrow[\substack{q_{\parallel} \to 0 \\ \xi \to 0}]{} & \frac{2}{\pi} \frac{q_{1,2}} {q_{LO}} \frac{e^{i q_{LO} L} - e^{\pm i q_{PhP,z} L}}{e^{i q_{LO} L} - e^{-i q_{LO} L}} \left( \varepsilon_{st} - \varepsilon_{\infty} \right) ,\\
\label{Seqn:Omega-minus-relation-LTSPhP}
\frac{\Omega_{-,\pm}}{i q_{\parallel} } & \xrightarrow[\substack{q_{\parallel} \to 0 \\ \xi \to 0}]{} &\frac{2}{\pi} \frac{q_{1,2}} {q_{LO}} \frac{e^{-i q_{LO} L} - e^{\pm i q_{PhP,z} L}}{e^{i q_{LO} L} - e^{-i q_{LO} L}}  \left( \varepsilon_{st} - \varepsilon_{\infty} \right) ,
\end{eqnarray} 
with $q_{LO} = \omega_{LO}/\beta_{L}$ is LO phonon wave vector. These terms modify the relations between $r_{pp}^{ij}$, $t_{pp}^{ij}$ as
\begin{equation}
\label{Seqn:rpp-relation-LTSPhP}
t_{pp}^{12} = \frac{r_{pp}^{12}+1}{\varepsilon_{PhP}}, \quad t_{pp}^{21} = \varepsilon_{PhP} \left( r_{pp}^{21}+1 \right), \quad r_{pp}^{21} = r_{pp}^{23} \neq - r_{pp}^{12}.
\end{equation} 
As a result, the dominant component of Fresnel reflection coefficient at the long wavelength limit and small frequency is
\begin{eqnarray}
\label{Seqn:rpp-LTSPhP-finite}
\mathbf{Re} \left( \left( r_{pp} (\mathbf{q}_{\parallel},i \xi) \right)^2 \right) \xrightarrow[\substack{q_{\parallel}\to 0 \\ \xi \to 0}]{L \to 0} \dfrac{1 - \left(\frac{\pi \varepsilon_{st} q_{LO}}{2 \left( \varepsilon_{st} - \varepsilon_{\infty} \right) q_{1,2}}\right)^2 \cot^2 \left( \frac{q_{LO} L}{2} \right) }{\left[ 1 + \left(\frac{\pi \varepsilon_{st} q_{LO}}{2 \left( \varepsilon_{st} - \varepsilon_{\infty} \right) q_{1,2}}\right)^2 \cot^2 \left( \frac{q_{LO} L}{2} \right) \right]^2} \equiv \mathcal{R}(L).
\end{eqnarray}
Thus, the quantum and thermal Casimir pressure are
\begin{eqnarray}
    \label{Seqn:Casimir-quantum-LTSPhP-finite}
P_{qm} (D \gg L) & \approx &  - \mathcal{R} (L)  \frac{\pi^2 \hbar c}{240 D^4},\\
    \label{Seqn:Casimir-thermal-LTSPhP-finite}
P_{th} (D \gg L) & \approx &  - \mathcal{R} (L)\frac{\zeta(3) k_B T}{8 \pi D^3}.
\end{eqnarray}
The function $\mathcal{R} (L)$ reaches its maximum $\mathcal{R} (L) = 1$ when $\cot \left( \frac{q_{LO} L}{2} \right) = 0$, it becomes negative when $\left| \cot \left( \frac{q_{LO} L}{2} \right) \right| > \frac{2 \left( \varepsilon_{st} - \varepsilon_{\infty} \right) q_{1,2}}{\pi \varepsilon_{st} q_{LO}}$, and it reaches its minimum $\mathcal{R} (L) = -0.125$ when $\cot \left( \frac{q_{LO} L}{2} \right) = \pm \frac{2 \sqrt{3} \left( \varepsilon_{st} - \varepsilon_{\infty} \right) q_{1,2}}{\pi \varepsilon_{st} q_{LO}}$. As observed from the numerical calculations, the Casimir pressure changes its sign and exhibits ``resonance''-like features when varying the piezoelectric plate 's thickness $L$. This is due to the appearance of the dominant term \eqref{Seqn:rpp-LTSPhP-finite} is result from the interplay between the cavity mode of LO phonons and the hybrid LT-SPhP modes. Here contribution of s-polarization is the same as SPhP, which is $D^{-6}$ and $D^{-5}$ in quantum and thermal limit, then be dominated by stronger scaling law of p-polarization.

\section{\label{appE}The Casimir force between piezoelectric materials at different temperatures}

For zero temperature, the Matsubara summation given in Eq. (10) is replaced by integral over imaginary frequencies 
\begin{equation}
\label{Seqn:small-temp}
\lim\limits_{T \to 0}k_B T {\sum_{n=0}^{\infty}}^{\prime} f(\xi_n) = \lim\limits_{T \to 0}k_B T \sum_{n=1}^{\infty} f(\xi_n) = \frac{\hbar}{2 \pi} \int_0^{\infty} f(\xi) d \xi .
\end{equation}
As the result, we obtained quantum Casimir pressure as $P_{qm} (D) \propto D^{-4}$. On the other hands, for extremely large temperature, only $n=0$ term in the Matsubara summation given in Eq. (10) is the most dominant one 
\begin{equation}
\label{Seqn:large-temp}
\lim\limits_{T \to \infty }k_B T {\sum_{n=0}^{\infty}}^{\prime} f(\xi_n) \to \frac{1}{2} k_B T f(0).
\end{equation}
As the result, we obtain so-called thermal Casimir pressure as $P_{th} (D,T) \propto D^{-3}$. For small finite temperature (about room temperature or less), the Matsubara summation given in Eq. (10) can be approximated as
\begin{eqnarray}
\label{Seqn:approx-temp}
&& k_B T {\sum_{n=0}^{\infty}}^{\prime} f(\xi_n) = k_B T {\sum_{n=1}^{\infty}}^{\prime} f(\xi_n) + \frac{1}{2} k_B T f(0) \approx \frac{\hbar}{2 \pi} \int_0^{\infty} f(\kappa) d \kappa + \frac{1}{2} k_B T f(0) \nonumber\\
\Rightarrow && P(D,T) \approx P_{qm} (D) + P_{th} (D,T).
\end{eqnarray}
The validity of this approximation is indicated by the following Fig. \ref{fig:S1}.

The Casimir pressure between two semi-infinite SiC plates supporting SPhPs or LT-SPhPs at different temperatures is shown in Fig. \ref{fig:S1}.

\begin{figure}[H]
\begin{center}
\includegraphics[width = 0.7 \columnwidth]{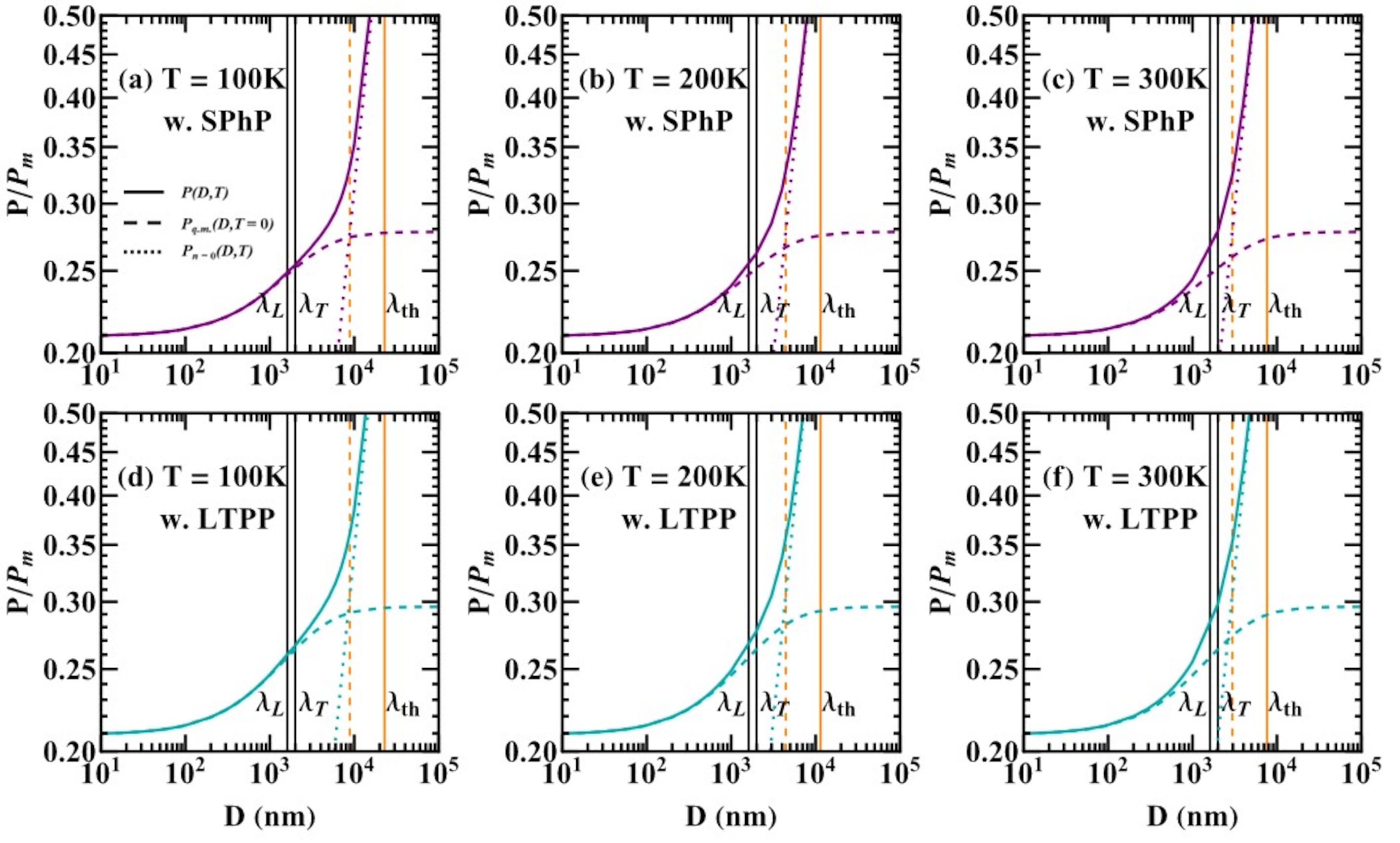}
\caption{\label{fig:S1}Casimir pressure normalized by the perfect metal limit $P_m$ between two semi-infinite SiC plates with SPhPs modes at (a) $T=100$ K, (b) $T=200$ K, (c) $T=300$ K. Casimir pressure normalized by the perfect metal limit $P_m$ between two semi-infinite SiC plates with LT-SPhPs modes at (c) $T=100$ K, (d) $T=200$ K, (e) $T=300$ K. Solid lines correspond to the pressure calculated using Eq. (10) in the main text, dashed lines correspond to the quantum mechanical limit $(T = 0)$, dotted lines correspond to the thermal limit $(T \gg \hbar c/k_B D)$. This indicates the validity of the approximation \eqref{Seqn:approx-temp}. The quantum-to-thermal transition distance $D_{th}$ is determined as the crossing between the quantum mechanical limit ($T = 0$, dashed lines) and the thermal limit ($T \gg \hbar c/k_B D$, dotted lines) and then be expressed as dotted vertical lines.}
\end{center}
\end{figure}

\begin{figure}[H]
\begin{center}
\includegraphics[width = 0.7 \columnwidth]{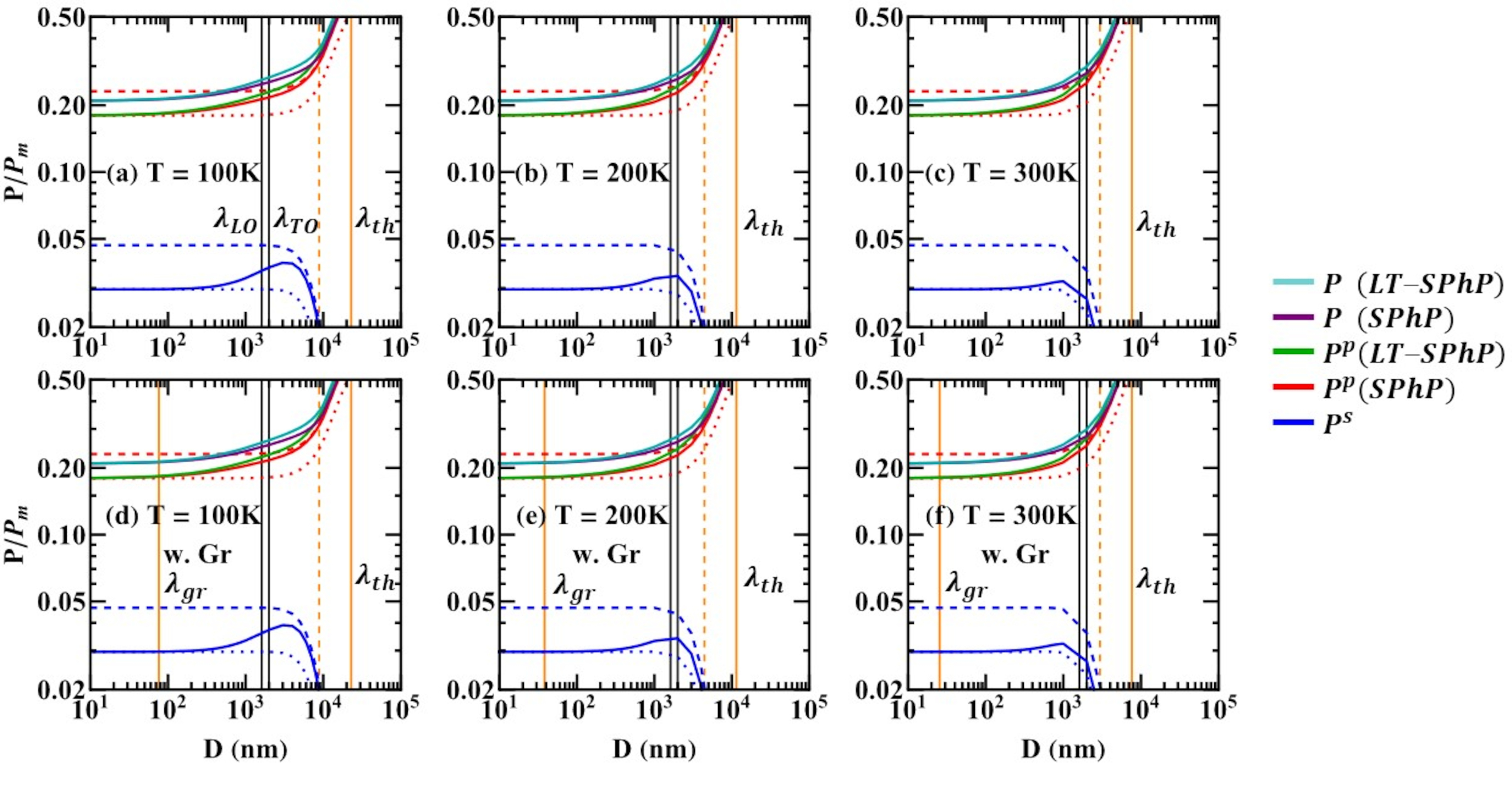}
\caption{\label{fig:S2} Casimir pressure normalized by the perfect metal limit $P_m$ between two semi-infinite SiC plates modes at (a) $T=100$ K, (b) $T=200$ K, (c) $T=300$ K. Casimir pressure normalized by the perfect metal limit $P_m$ between two semi-infinite SiC plates with graphene monolayers at (c) $T=100$ K, (d) $T=200$ K, (e) $T=300$ K.}
\end{center}
\end{figure}

The Casimir pressure between two semi-infinite SiC plates covered by graphene monolayers supporting SPhPs or LT-SPhPs at different temperatures is shown in Fig. \ref{fig:S2}.

\begin{figure}[H]
\begin{center}
\includegraphics[width = 0.7 \columnwidth]{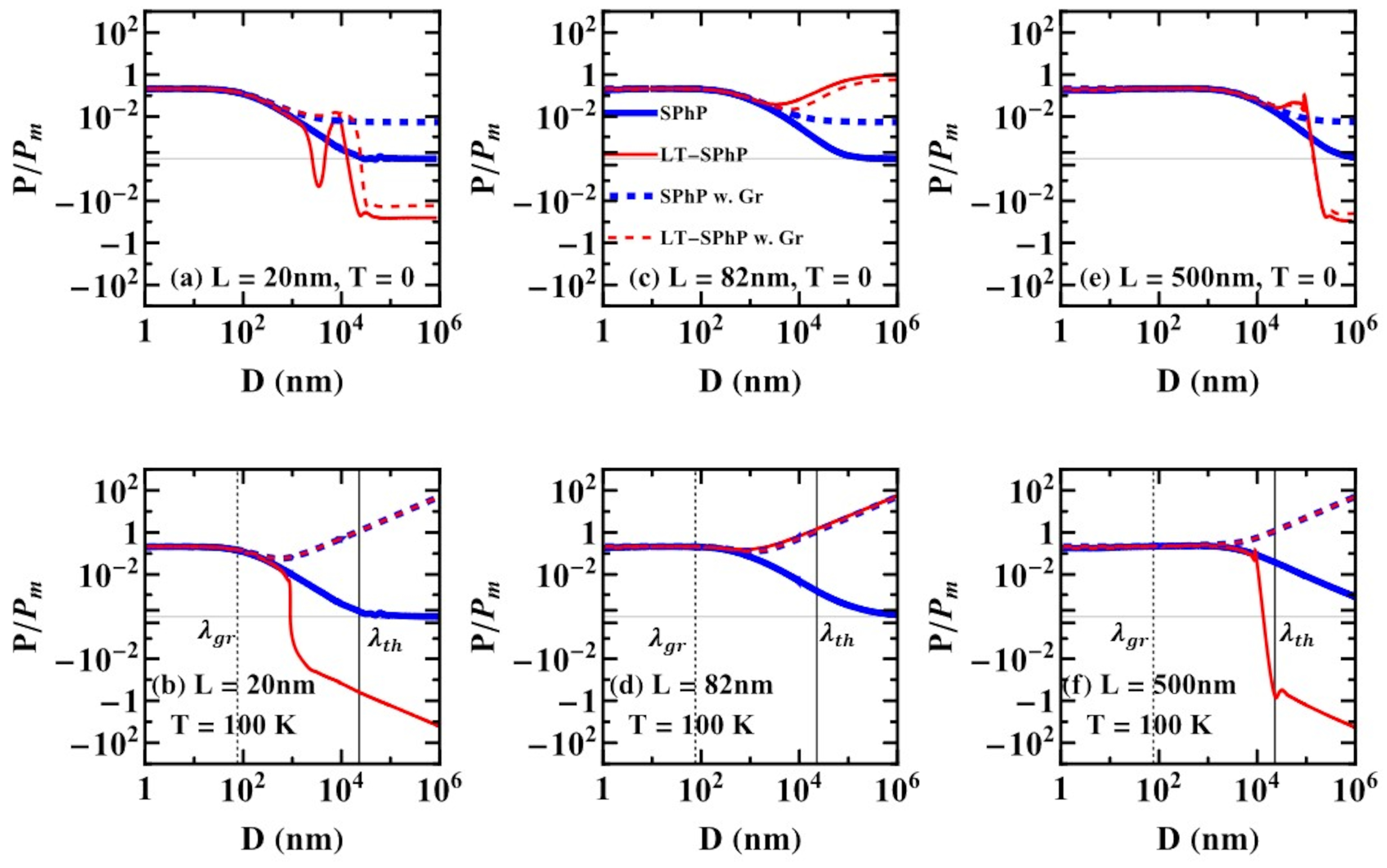}
\caption{\label{fig:S3} Symmetric log scale representation for Casimir pressure normalized by the perfect metal limit $P_m$ between two SiC plates modes at $T=0$ K and with  (a) $L=20$ nm, (b)$L=82$ nm, (c) $L=500$ nm.  Symmetric log scale representation for Casimir pressure normalized by the perfect metal limit $P_m$ between two SiC plates modes at $T=100$ K and with  (a) $L=20$ nm, (b)$L=82$ nm, (c) $L=500$ nm. }
\end{center}
\end{figure}

The Casimir pressure between two SiC plates supporting SPhPs or LT-SPhPs with different thickness is shown in Fig. \ref{fig:S3} in symmetric log-log scale. This log-log scale enables a better analysis of the scaling laws of the Casimir interaction between the plates.

%


\end{document}